\documentclass[10pt]{article}
\usepackage[utf8]{inputenc}
\usepackage{authblk}
\usepackage{amsmath}
\usepackage{graphicx}
\usepackage{setspace}
\usepackage{xcolor}
\usepackage[margin=1in]{geometry}

\usepackage[backend=biber,style=numeric,sorting=none]{biblatex}
\bibliography{pmaca1refs.bib}
\title{Physical mechanisms affecting critical angle for nanopatterning in irradiated thin films: I. A composite model}
\author{Tyler Evans \& Scott Norris}
\affil{Department of Mathematics, Southern Methodist University, Dallas, TX 75275, United States of America}

\begin{document}

\maketitle

\begin{abstract}
Ion-beam irradiation of an amorphizable material such as Si or Ge may lead to spontaneous pattern formation, rather than flat surfaces, for irradiation beyond some critical angle against the surface normal. It is observed experimentally that this critical angle varies according to many factors, including beam energy, ion species and target material. However, most prevailing theoretical analyses predict a critical angle of $45^{\circ}$ independent of energy, ion and target, disagreeing with experiment. Previous work on this topic \cite{moreno-barrado-etal-PRB-2015} has suggested that inhomogeneous bulk stress may modify critical angle selection. However, this analysis was done using a differently-motivated stress tensor belonging to the ``effective body force" class of models, rather than the ``stress-free strain rate" class of models. A specifically angle-independent isotropic stress component, which has been experimentally observed and theoretically studied, was also neglected.  In this first part of a set of papers, we consider a composite model of stress-free strain and isotropic swelling with a generalized treatment of stress modification along idealized ion tracks. We obtain a highly-general linear stability result with a careful treatment of arbitrary depth-dependence profiles for each of the stress-free strain-rate tensor, a source of deviatoric stress modification, and isotropic swelling, a source of isotropic stress. We compare our theoretical results with experimental measurements of angle-dependent deviatoric and isotropic stresses for 250eV Ar$^+$ on Si. Our analysis suggests that the presence of angle-independent isotropic stress and the relationship between the amorphous-crystalline and free interfaces may be strong contributors to critical angle selection, while the influence of inhomogeneous stress modification is seemingly non-existent in the idealized case of diagonally-translated interface and stress generated entirely along a thin, down-beam ion track. We also consider an opposing idealization: that of interfaces defined by vertical translation, with stress modification along the vertical regardless of beam orientation. The unacceptable variability in predictions resulting from these two idealizations, both of which have appeared in recent analyses, prompts modeling refinements discussed in subsequent papers in this set. These refinements include the relationship between interfaces and a more sophisticated treatment of the inhomogeneous stress field.
\end{abstract}

\tableofcontents

%\doublespacing

\section{Introduction}
It has been known since at least 1960's that some materials may be irradiated by an ion beam in an energy range of 100eV to 10keV in order to produce nanoscale patterns of various kinds \cite{navez-etal-1962}. Examples include ripples, hexagonal arrays of dots, and continuous transitions between the two \cite{facsko-etal-SCIENCE-1999,frost-etal-APA-2008}. These structures can range in scale from approximately 5-20nm \cite{facsko-etal-SCIENCE-1999}. Because ion beams are already ubiquitous  in industrial settings, these early findings inspired great interest in developing a comprehensive theory of nanostructuring. It was thought that the tendency for surfaces to spontaneously self-organize into coherent patterns could be controlled and applied in a mass-manufacturing context, facilitating the production of cheap and high-quality nano- and meta- materials. Rather than precisely engineering a structure part-by-part, these materials could be ``grown" through a ``bottom-up" approach, requiring only coarse control of the system at scales many orders of magnitude greater than that of the resulting nanostructures. However, despite decades of effort, such a comprehensive theory has remained elusive \cite{chan-chason-JAP-2007,munoz-garcia-etal-MSER-2014,NorrisAziz_predictivemodel}, and the dream of ``bottom-up" nano-engineering has gone (lamentably) unrealized at the time of writing.

The lack of a universal theory aside, some things have come to be understood. Specifically, it is known that ion-induced erosion rates are not constant across a given film topography: the bottoms of pits may erode faster than the peaks on the surface, which can result in the well-known Bradley-Harper instability \cite{sigmund-PR-1969,sigmund-JMS-1973,bradley-harper-JVST-1988,makeev-etal-NIMB-2002}. The orientation and wavelength of ripples caused by ion-bombardment exhibit dependence on the incidence angle of the beam as well as its energy, enabling some small measure of empirical controllability \cite{carter-vishnyakov-PRB-1996,moseler-etal-SCIENCE-2005,davidovitch-aziz-brenner-JPCM-2009,madi-etal-PRL-2011}, although this is insufficient to achieve manufacturing applications. It has also proven useful to consider mechanisms acting on two time scales: the ``prompt regime", on the order of $\sim10^{-9}$ seconds, and the ``gradual regime", on the order of $\sim10^2$ seconds or longer. Within the prompt regime, erosive and redistributive effects dominate, and the primary focus is on the free interface. A recent, integrated approach to handling both erosion and redistribution simultaneously is the crater function framework \cite{kalyanasundaram-etal-APL-2008,kalyanasundaram-etal-JPCM-2009, norris-etal-2009-JPCM,norris-etal-NCOMM-2011,harrison-bradley-PRB-2014}, a long-wave approximation of the surface that utilizes simulation data to collect information about the ``craters" caused by individual ion impacts, which may then be used to study aggregate behavior. A number of variations are currently in use.

It has also been established that some materials, upon sustained exposure to ion bombardment, may develop a thin, amorphous layer near the surface, which can be appropriately modeled as a highly viscous fluid \cite{umbach-etal-PRL-2001}. It is within this slow-moving film that gradual-regime effects act. That the gradual regime is \textit{slow} should not be taken to imply that it is less important in determining surface evolution than the prompt regime phenomena. Indeed, within the gradual regime, such mechanisms as stress buildup and relaxation, surface diffusion, and viscous flow may occur \cite{volkert-JAP-1991,snoeks-etal-JAP-1995,brongersma-etal-JAP-2000,van-dillen-etal-APL-2001-colloidal-ellipsoids,van-dillen-etal-APL-2003-colloidal-ellipsoids,van-dillen-etal-PRB-2005-viscoelastic-model,mayr-averback-PRB-2005,otani-etal-JAP-2006,chan-chason-JVSTA-2008,madi-etal-2008-PRL,madi-thesis-2011,madi-aziz-ASS-2012,ishii-etal-JMR-2014}, and it turns out that viscous relaxation, a gradual-regime mechanism, is the most likely source of regularization even in nonlinear surface evolution. This was, even recently, a matter of debate, as the original, erosive-redistributive theory used crystallographic surface diffusion instead. Such surface diffusion should be temperature-dependent, which has now been contradicted by at least two experimental-theoretical collaborations \cite{umbach-etal-PRL-2001, norris-etal-SREP-2017}.

Many other mechanisms have been suggested to act in this regime to influence pattern formation, such as effective body forces \cite{castro-cuerno-ASS-2012,munoz-garcia-etal-PRB-2019}, anisotropic plastic flow \cite{norris-PRB-2012-linear-viscous}, and isotropic swelling \cite{Swenson_2018}. Because these mechanisms are slower, smaller in length-scale and active \textit{within} the amorphous bulk, they may be more difficult to experimentally observe when compared with prompt-regime surface phenomena. Nonetheless, they are increasingly viewed as important to fully understand, and cataloging all such mechanisms remains a rich source of theoretical-experimental collaboration, especially as evidence mounts that the erosion may be a weaker contributor to nano-scale pattern formation than was previously thought in at least some energetic regimes \cite{norris-etal-NCOMM-2011, norris-etal-SREP-2017}.

Despite advances in modeling irradiated surfaces, no unifying theory exists. However, linear stability analysis in Fourier modes has proven to be a very useful tool in bridging theory and experiment, as experimental tools such as GISAXS naturally produce wavelength data for surface ripples. It is also simple to experimentally study the critical beam-angle at which the irradiated surface transitions from flat to patterned, and many such studies exist; for example, \cite{madi-aziz-ASS-2012}. Conveniently, both critical angle and wavelength predictions naturally emerge from such modal analysis. Any candidate for a unifying theory of nano-scale pattern formation must therefore, as a minimal condition, succeed in correctly predicting the experimentally-observed critical angle and the wavelengths across a variety of systems, including the possibility that some systems may form no patterns at all \cite{Teichmann2013,hofsass-bobes-zhang-JAP-2016}

Seemingly an indictment of the present state of theory, it is not yet even understood which mechanisms determine the critical beam-angle for which nanopatterning begins \cite{munoz-garcia-etal-MSER-2014,NorrisAziz_predictivemodel}. Experimental results for 250eV-1keV Ar$^+$ on Si yield $\theta_c \approx 45^{\circ}$ \cite{madi-etal-2008-PRL}, whereas heavier ions such as Xe$^+$ and Kr$^+$ on Si at similar energies result in $\theta_c \approx 60^{\circ}$ \cite{perkinsonthesis2017}. Other values have been obtained experimentally for different ion, energy and target combinations. In contrast, most of the prevailing theoretical analyses of the hydrodynamic type have all predicted $\theta_c = 45^{\circ}$ as an apparently-universal feature \cite{castro-cuerno-ASS-2012, castro-etal-PRB-2012,norris-PRB-2012-linear-viscous}, which is clearly wrong. Such a discrepancy between theory and experiment requires an amended theory. 

It is noteworthy that a variant of the crater function framework purports to have explained critical angle selection through the inclusion of an empirical ``curvature coefficient" that takes into account local height variations and modifies the kinematic condition at the free interface \cite{hofsass-bobes-zhang-JAP-2016}. Although this approach appears to lead to good agreement across several experimental systems, concerns have been raised with the physical justification for this coefficient \cite{NorrisAziz_predictivemodel}. One way or the other, as the crater function framework is a prompt-regime family of models, and inherently a long-wave approximation, it can \textit{never} reliably produce wavelength predictions, even if one of its variants \textit{does} correctly determine the critical angle on an unobjectionable basis. This is because critical angle selection is, itself, an inherently long-wave instability (a so-called ``Type-II" instability \cite{NorrisAziz_predictivemodel}. The disadvantages of the crater function framework due to its restriction to the long-wave limit disappear in the hydrodynamic family of models, which readily produce wavelength predictions from linear stability analyses conducted without any restrictions on wavenumber. It is therefore of practical value that the gradual-regime hydrodynamic and prompt-regime crater function families be unified into a single framework if possible. Indeed, if it is true that crater functions correctly predict critical angles, then it will be necessary to adapt those insights into the hydrodynamic class of models in order to make further progress towards predicting wavelengths and advancing the project of experimental-theoretical alignment.

In this direction, there has been recent consideration of some avenues for the introduction of material-specificity within the hydrodynamic family of models: in particular, depth-dependence in combination with phenomenological modeling of ion-induced stress is purported to lead to good fits with MD simulation data, both in the linear and nonlinear regimes \cite{moreno-barrado-etal-PRB-2015, munoz-garcia-etal-PRB-2019}. On the other hand, because there are so many competing mechanisms at work within nanostructure-forming systems, development of theory is highly nuanced and must be approached cautiously. Indeed, while a proposed mechanism, or its mathematical form as appears in, e.g., an evolution equation, may apparently lead to good predictions within one regime, it is possible that this is not due to having modeled the correct physics, but, rather, having found a mechanism that has the same behavior as the correct physics within the selected regime, but not others. In fact, this has already occurred, as we mentioned earlier: there was a brief controversy surrounding the correct regularization term, with theoretical work having variously used, on phenomenological bases, thermally-activated surface diffusion, surface-confined viscous flow, and an athermal effective surface diffusion activated by erosion. These mechanisms exhibit similar regularization behavior, scaling as $\sim q^4$ for small wavenumber $q$, while scaling quite differently from $q \to \infty$. Recently, experimental work has substantially bolstered the hypothesis that the surface-confined viscous flow model of regularization appears to be the most accurate based on analysis of GISAXS data \cite{norris-etal-SREP-2017}.

In the present set of papers, we revisit the anisotropic plastic flow model belonging to the hydrodynamic class, which has previously demonstrated a high level of agreement between experimental wavelength measurements and theoretical predictions for the 250eV Ar$^+$ irradiation of Si using only experimentally-determined quantities and no free parameters \cite{norris-PRB-2012-linear-viscous} and simple, commonly-used assumptions about interfacial geometry. In this first part of the series, we generalize this model to incorporate two additional features. First, we develop the capability of inserting arbitrary spatial dependence for the stress tensor associated with anisotropic plastic flow, as a substantial generalization of \cite{moreno-barrado-etal-PRB-2015,munoz-garcia-etal-PRB-2019}, which correctly noted that one should not anticipate that stress development would be uniform throughout the film. Second, we allow for a spatially-varying rate of isotropic swelling due to, e.g., radiation damage in the style of \cite{Swenson_2018}. Contrary to the results of \cite{moreno-barrado-etal-PRB-2015, munoz-garcia-etal-PRB-2019}, we find that our approach predicts \textit{no effect} on critical angle selection for depth-dependence induced by modified stress entirely along an ion-track where certain assumptions have been made about the relationship between the free and amorphous-crystalline interfaces. We acknowledge the subtle differences between our model and that of \cite{moreno-barrado-etal-PRB-2015} in the Discussion section and in the Appendix. Alternatively, and motivated by the experimental results of \cite{perkinsonthesis2017}, we propose that a combination of isotropic swelling, a careful consideration of the upper-lower interface relation, and stress modification outside of the idealization of a vanishingly-thin ion track may help explain the observed nonuniversality of $\theta_c$, bringing theory and experiment closer to alignment.

\section{Model}
\begin{figure}[h!]
	\centering
	\includegraphics[totalheight=4.5cm]{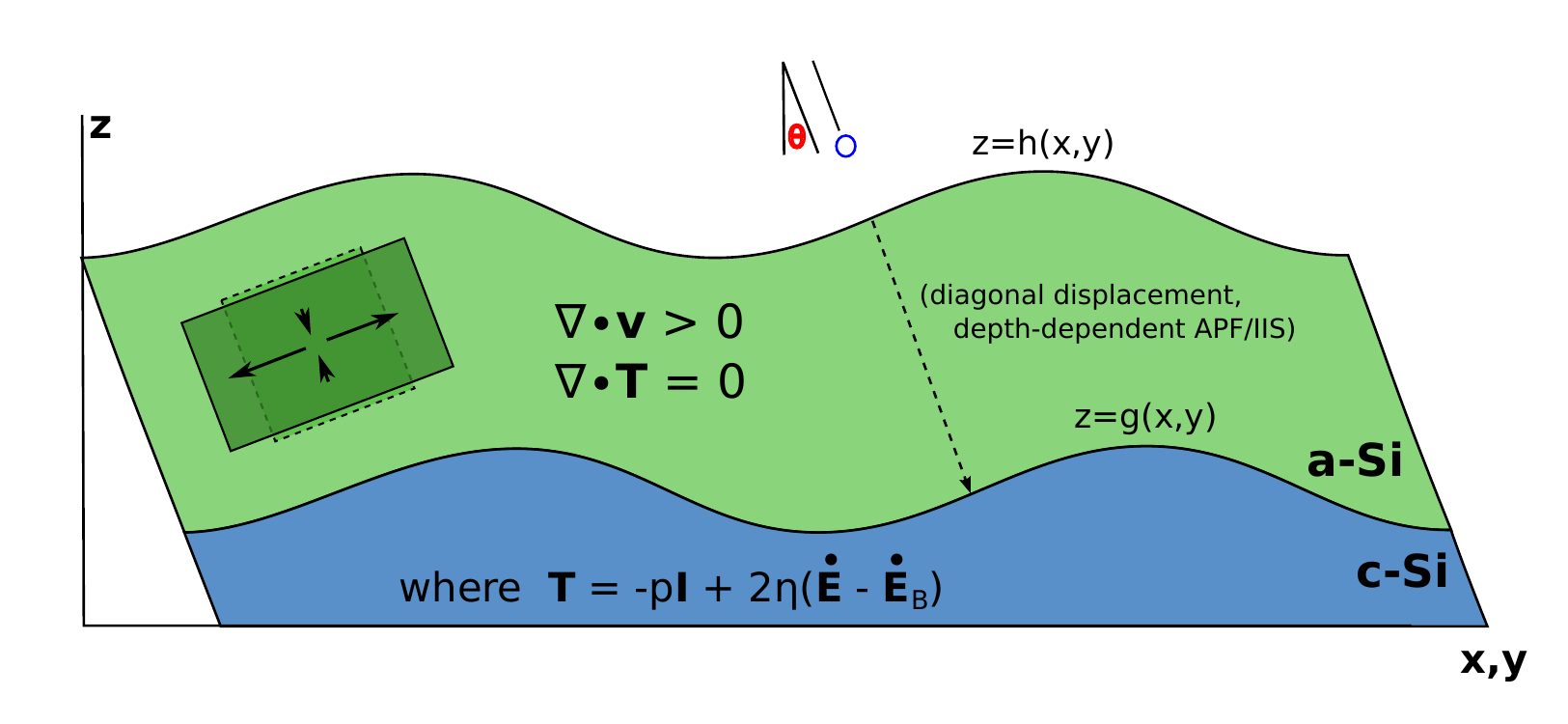}
	\caption{Schematic depicting ion bombardment at an incidence angle of $\theta$ and stress induced in the thin film by ion implantation.  Note that for off-normal incidence, the bottom boundary $z=g$ may not vertically align with the top boundary $z=h$. Here, we will make the simplifying assumption that all spatial variation occurs along the ``down-beam" direction.}
	\label{fig:schematic}
\end{figure}

\subsection{Physical description}
\paragraph{Anisotropic plastic flow (APF).}
By \textit{anisotropic plastic flow} (APF), we mean flow that has directionality (anisotropy) and which does not exhibit a stress response to the strain rate (plasticity). Typically, fluids behave according to the viscous stress tensor, which relates the amount of stress in the film to the rate at which the film deforms (``strains"). In contrast with other models which have suggested that the ion beam acts upon the film by exerting an ``effective body force" or adding to the stress tensor \cite{castro-cuerno-ASS-2012,moreno-barrado-etal-PRB-2015,munoz-garcia-etal-PRB-2019}, the anisotropic plastic flow model posits that the action of the beam is to \textit{deduct} a stress-free strain rate from the viscous stress tensor \cite{norris-PRB-2012-linear-viscous,van-dillen-etal-APL-2001-colloidal-ellipsoids,van-dillen-etal-APL-2003-colloidal-ellipsoids,van-dillen-etal-PRB-2005-viscoelastic-model}. That is: the beam, through repeated ion impacts and displacement of material, allows a greater strain rate for a given level of stress within the film than would otherwise occur, inducing a tendency for down-beam compression and cross-beam elongation. For this reason, the effect of anisotropic plastic flow is sometimes characterized as ``pancake strain" \cite{perkinsonthesis2017}. Although the notion of anisotropic plastic flow was originally used in the context of high energy irradiation in the MeV range \cite{trinkaus-NIMB-1998-viscoelastic,trinkaus-ryazanov-PRL-1995-viscoelastic,brongersma-etal-JAP-2000} , with localized melting as the mechanism underlying the effect, it was later observed in the keV range on a phenomenological basis, with the underlying mechanism currently unknown \cite{van-dillen-etal-APL-2001-colloidal-ellipsoids,van-dillen-etal-APL-2003-colloidal-ellipsoids,van-dillen-etal-PRB-2005-viscoelastic-model}. However, the use of the stress tensor associated with MeV anisotropic plastic flow has, on a purely phenomenological basis, led to surprising agreement with experimental resuls in both the linear and nonlinear regimes \cite{norris-PRB-2012-linear-viscous,george-etal-JAP-2010}.

We will consider a model of anisotropic plastic flow based on that of \cite{norris-PRB-2012-linear-viscous}, which is fundamentally that of Stokes flow modified by the addition of a stress-free strain rate to the viscous stress tensor. In \cite{norris-PRB-2012-linear-viscous}, this stress-free strain rate is a constant throughout the film depth, leading to its appearance in the governing equations only at the stress balance at the upper interface of the film. The stress tensor used is
\begin{equation}
	\textbf{T} = -p\textbf{I} + 2\eta\left(\dot{\textbf{E}} - fA\textbf{D}(\theta)\right),
\end{equation}
where $fA\textbf{D}(\theta)$ is the stress-free strain rate, anisotropically permitting more strain than would be possible in a typical Newtonian fluid. With $fA\textbf{D}(\theta) \to 0$, this reduces to the standard viscous stress tensor, and any subsequent linear stability analysis would describe only the Orchard mechanism of viscous surface relaxation \cite{orchard-ASR-1962}.

\paragraph{Ion-induced swelling (IIS).}
Ion-induced swelling (IIS) appears through the equation
\begin{equation}
	\frac{\partial \Delta}{\partial t} + \vec{v}\cdot \nabla \Delta = \alpha(z;g,h),
\end{equation}
where $\Delta$ tracks the amount of ``volumization" or ``swelling" sustained by a parcel of matter as it dwells within the film prior to sputtering away, acquired at the rate $\alpha(z;g,h)$ which has units $\frac{1}{\text{s}}$. We note that this is a generalization of the equation
\begin{equation}
	\frac{\partial a}{\partial t} + \vec{v}\cdot \nabla a = 1,
\end{equation}
which was used in the work of \cite{Swenson_2018,evans-norris-JPCM-2022}. When we take $\Delta = \alpha a$, with $\alpha$ constant, our equation in $\Delta$ reduces to the above. We then take as the equation of state
\begin{equation}
	\rho(\Delta) = \frac{\rho^*}{1+\Delta}
\end{equation}
which assigns to each parcel of matter a density that is penalized due to the damage that has been sustained there over time (hence swelling). Elsewhere \cite{Swenson_2018,evans-norris-JPCM-2022}, such models have been shown to increase the linear stability of thin-film surfaces against topographical perturbations, increasing the critical angle $\theta_c$.

\paragraph{Spatial variation in mechanism intensity: need for generalization.}
It has been discussed elsewhere \cite{moreno-barrado-etal-PRB-2015} that a linearly depth-dependent beam-induced stress tensor may alter critical angle selection. In the present work, we also consider bulk mechanisms whose strength varies with the depth for each of two different depth-dependence profiles along with the two bulk mechanisms discussed above, anisotropic plastic flow and isotropic swelling. The question of how each of these mechanisms vary with depth is linked to the question of the fundamental physics underlying each of the two mechanisms under consideration. 

As discussed above, the stress tensor $fA\textbf{D}(\theta)$, used here and in other work is used on a phenomenological basis and borrowed from the electronic stopping regime observed in the Ar$^+ \to$ Si system above $\sim 100$ keV for which good agreement between theory and experiment has been attained. It is unexpected that the same stress tensor would produce good agreement between theory and experiment at such low energies, which are well-within the nuclear stopping regime, as the stress tensor emerges from a ``melt-cycle" model caused by displacement spikes, and their subsequent thermalization, local to the track of the incoming ion \cite{wesch-wendler-book-2016}. Due to the inherent uncertainties surrounding the underlying physics contributing to each of the two bulk-active mechanisms considered here, we will conduct our analysis in the fullest possible generality with respect to these depth-dependence profiles, permitting a broader, if more speculative, exploration of the implications of each hypothesis. There exist at least two possibilities that may explain this phenomenon, each of which will have different implications for the correct theoretical modeling of beam-induced stress.

One possibility is that APF is driven through \textit{power deposition by nuclear and electronic stopping.} Although for energies below about 10keV energy deposition via electronic stopping is seemingly negligible in comparison with that of nuclear stopping \cite{ziegler-biersack-littmark-1985-SRIM,ziberi-etal-APL-2008,liedke-thesis-2011}, it is possible that \textit{some} atomic displacements still occur along the ion track which qualitatively resemble the thermalization within the electronic stopping regime \cite{klaumunzer-etal-REDS-1989}. In this case, we would expect that the depth-dependence of anisotropic plastic flow should be that of a decaying exponential about the ion-tracks, and a depth-dependence through the bulk should be generated accordingly. It has elsewhere been suggested that, at least within some energetic ranges \cite{steinbach-etal-PRB-2011}, swelling (implied to be isotropic) is associated with energetic deposition by nuclear stopping. If this is true within the regime of dominant nuclear stopping as well, then we would expect a depth-profile for the isotropic swelling rate that resembles a Gaussian in the downbeam direction or along the ion-track, approximately following the shape of the collision cascade wherein the majority of the incoming ion's energy is lost due to nuclear stopping.

A second possibility is that APF is driven by \textit{artifacts of redistribution and defect dynamics.} It has been discussed elsewhere \cite{norris-PRB-2012-linear-viscous} that the apparent good agreement between experimental observations and the $fAD(\theta)$ stress tensor could be due to anisotropic plastic flow in the nuclear stopping regime being, in actuality, the same redistributive phenomenon considered in the crater function framework family of models \cite{kalyanasundaram-etal-APL-2008,kalyanasundaram-etal-JPCM-2009, norris-etal-2009-JPCM,norris-etal-NCOMM-2011,harrison-bradley-PRB-2014} and which had previously been studied by \cite{carter-vishnyakov-PRB-1996}. Observations in favor of this view include the work of \cite{lopezcazalilla-et-al-2018} in the negligible sputtering regime which showed topographical changes to a thin Si film under Ar$^+$ irradiation at 30eV (below the sputter threshold for Ar$^+$ on Si of 33eV) analogous to those observed in 250eV irradiation of the same. It is well-understood that numerous atomic displacements occur within irradiated films due to intra-film collision cascades, and the dynamics of the resulting defects are believed to be responsible for the amorphization process itself \cite{wesch-wendler-book-2016}. If both anisotropic plastic flow and isotropic swelling are fundamentally due to rearrangement of material within the film as by, e.g., void formation within the collision cascade and the relaxation of the resulting interstitials outside of the collision cascade relaxing to induce an anisotropic stress, we might expect both mechanisms to exhibit the same depth dependence, being strongest at the center of the collision cascade and weakening outward with the deposition of energy via nuclear stopping.

\subsection{Governing equations}
We will study a composite model including APF and IIS. First, we have the equations associated with momentum conservation in the bulk,
\begin{equation}
	\nabla \cdot \textbf{T} = 0,
\end{equation}
which is simply the differential form of bulk momentum conservation in the well-known limit of Stokes flow (i.e., large Reynolds number or creeping, thin-film flow), where we make use, on a phenomenological basis, of the stress tensor associated with anisotropic plastic flow in the MeV regime \cite{trinkaus-NIMB-1998-viscoelastic,trinkaus-ryazanov-PRL-1995-viscoelastic,van-dillen-etal-PRB-2005-viscoelastic-model,norris-PRB-2012-linear-viscous,wesch-wendler-book-2016}:
\begin{equation}
	\begin{gathered}
		\textbf{T} = - p\textbf{I} + 2\eta\{\dot{\textbf{E}} - \dot{\textbf{E}}_b\} \\
		\dot{\textbf{E}} = \frac{1}{2}\left( \nabla \vec{v} + \nabla \vec{v}^T \right) \\
		\dot{\textbf{E}}_b = fA\tau(z;g,h) \textbf{D}(\theta) \\
		\textbf{D}(\theta) = 
		\begin{bmatrix} 
			\frac{3}{2}\cos(2\theta) - \frac{1}{2} & 0 & \frac{3}{2}\sin(2\theta) \\
			0 & 1 & 0 \\
			\frac{3}{2}\sin(2\theta) & 0 & -\frac{3}{2}\cos(2\theta) - \frac{1}{2} \\
		\end{bmatrix}.
	\end{gathered}
\end{equation}
Considering mass conservation in the bulk, we have
\begin{equation}
	\begin{gathered}
		\frac{\partial \rho}{\partial t} + \nabla \cdot (\rho \vec{v}) = 0 \\
		\frac{\partial \Delta}{\partial t} + \vec{v}\cdot \nabla(\Delta) = \alpha(z;g,h) \\
		\rho = \frac{\rho^*}{1+\Delta},
	\end{gathered}
\end{equation}
where the first equation is simply the differential form of mass conservation, the second equation tracks the ion-induced damage incurred by the film as parcels of matter advect throughout the bulk, and the third is an equation of state relating local density $\rho$ to the accumulation of damage ($\Delta$) there and the initial crystalline density $\rho^*$. Elsewhere \cite{Swenson_2018,evans-norris-JPCM-2022}, this has been referred to as a form of ``quasi-incompressibility", in that density $\rho$ retains no explicit dependence on pressure $p$, and depends only on volumization $\Delta$. 
Note that in the case $\Delta = a\alpha$ with $\alpha = \text{constant}$, this equation reduces to that of \cite{Swenson_2018}. In principle, it is of course possible to have some other equation of state.

\noindent At the free upper interface, $z=h$, we have
\begin{equation}
	\begin{gathered}
		v_{I,h} = \vec{v}\cdot \hat{n} - V\frac{\rho^*}{\rho} \\
		\textbf{[T]} \cdot \hat{n} = -\gamma \kappa \hat{n}
	\end{gathered}
\end{equation}
where the first is a modified kinematic condition due to mass conservation at the upper interface, which takes into account the removal of material due to erosion. Considering at least the steady-state velocity due to sputtering, $V$, has turned out to be necessary for the isotropic swelling mechanism to attain a steady state; otherwise, there will be ``infinite volumization". For details of the derivation, see (cite Swenson-Norris). At the lower, amorphous-crystalline interface, $z=g$, we have
\begin{equation}
	\begin{gathered}
		\Delta = 0 \\
		\vec{v}\cdot \hat{t} = 0, \\
		\vec{v}\cdot \hat{n} = 0,
	\end{gathered}
\end{equation}
and these latter two equations are simply the imposition of no-slip and no-penetration conditions at the amorphous-crystalline boundary, which reduce identically to
\begin{equation}
\begin{gathered}
	u = 0 \\
	v = 0
\end{gathered}
\end{equation}
at $z=g$. It is intuitive that $\Delta = 0$ should be the case because as the film erodes downward, the material at the crystalline-amorphous interface should always be the newest, hence having sustained no instantaneous damage (yet) at any moment. In the above, $[\hspace{.25cm}]$ denotes the jump across the material interface. We convert to a moving frame traveling downward with the eroding surface at steady-state erosion velocity V; mathematically,
\begin{equation}
	\begin{gathered}
		h \to h-Vt \\
		g \to g-Vt \\
		v_{I,h} \to v_{I,h} - V(\hat{k}\cdot\hat{n}) \\
		z \to z-Vt \\
		\vec{v} \to \vec{v} - V\hat{k}.
	\end{gathered}
\end{equation}

\section{Analysis}
\subsection{Linear stability}
Here, we discuss the main calculations of the present work as a summary. For full details of the expansion and the systems of ordinary differential equations solved in the basic case, we refer the reader to the Appendix; the calculations are lengthy and do not, in themselves, contribute to the discussion. 
\paragraph{Formulation.} We construct a linearization of the previously-described governing equations about steady-state solutions in each of the bulk fields $\rho, \vec{v}, p, \Delta$ and the two interfaces $g$ and $h$ as
\begin{equation}
\begin{gathered}
	\rho(x,z,t) \to \rho_0(z) + \epsilon \rho_1(x,t) \\
	u(x,z,t) \to u_0(z) + \epsilon u_1(x,t) \\
	w(x,z,t) \to w_0(z) + \epsilon w_1(x,t) \\
	p(x,z,t) \to p_0(z) + \epsilon p_1(x,t) \\
	\Delta(x,z,t) \to \Delta_0(z) + \epsilon \Delta_1(x,t) \\
	h(x,t) \to h_0 + \epsilon h_1(x,t) \\
	g(x,t) \to g_0 + \epsilon g_1(x,t) \\
\end{gathered}
\end{equation}
where we have stripped out y-dependence (hence the bulk field $v$) in order to focus on the equations describing the evolution of the system in the projected downbeam direction. After collecting powers of $\epsilon$, we obtain the steady-state equations the linearized, O($\epsilon$) equations, which can be solved individually. The O($\epsilon$) equations will, in general, have dependence on the steady states. We may subsequently take each of $\rho_1,u_1,v_1,w_1,p_1,\Delta_1,h_1,g_1$ as, for example,
\begin{equation}
	\rho_1(x,t) = \tilde{\rho}_1\exp(\sigma t + ikx),
\end{equation}
which is referred to as an expansion in normal modes. Then, everywhere in the governing equations, we find that $\frac{\partial}{\partial t} \to \sigma$ and $\frac{\partial}{\partial x} \to ik$, which greatly simplifies the analysis and converts the system in the leading-order correction terms (O($\epsilon$)) into ordinary differential equations in independent variable $z$, the film depth in laboratory coordinates (the familiar Cartesian x-z). Solving this system yields $\sigma$, the dispersion relation, as a compatibility requirement for the linearized kinematic condition, from which we may obtain theoretical predictions of useful quantities, such as the most unstable wavenumber (leading to predictions of experimentally-observed wavelength) and bifurcation angle $\theta_c$. By $\theta_c$, we mean the critical beam angle at which surface perturbations transition from stable (decaying with time) to unstable (growing with time), a Type II bifurcation known to be associated with the long-wave limit \cite{NorrisAziz_predictivemodel}, hence small wavenumber $k$. 

Because we are primarily interested in $\theta_c$, we further perform a Taylor expansion in powers of wavenumber $k \approx 0$. The resulting system is still very difficult or impossible to solve analytically. Due to the nature of the analysis, we have a strong preference for a closed-form solution. A simplifying assumption leading to analytical tractability is the limit of small swelling rate, $\hat{\alpha} \approx 0$. Based on experimental estimates of deformation rates, we anticipate that this quantity should be small for low energies in the Ar$^+ \to$ Si system, especially the $\leq$ 10keV range of our immediate interest. As a result of our analysis, we obtain a general solution expressed entirely in terms of the depth-dependence profiles, which are left arbitrary.

\paragraph{Linearization of the intensity fields.} 
In the course of the linear stability analysis, we are naturally required to perform a linearization of the spatial-variation profiles associated with each bulk mechanism. In the present work, we will derive these profiles from angle-dependent intensity fields associated with the generation of each respective bulk mechanism along vanishingly-thin ion tracks. In the spirit of generality, we carry out our linear stability analysis for arbitrary depth-dependence profiles, which forms a reusable framework for later analyses. We will therefore consider expansions
\begin{equation}
\begin{gathered}
	\tau(z;...) = \tau_0(z;...) + \epsilon\tau_{\epsilon}(z;...) + O(\epsilon^2) \\
	\alpha_1(z;...) = \alpha_{10}(z;...) + \epsilon\alpha_{1\epsilon}(z;...) + O(\epsilon^2)
\end{gathered}
\end{equation}
in anticipation of these forms arising within the linear stability analysis. Here, the $(...)$ denotes any other quantities carried within the linearization. For example, linearizing as $g = 0 + \epsilon g_1$ and $h = h_0 + \epsilon h_1$ leads to
\begin{equation}
	\tau(z;g,h) = \tau(z;0,h_0) + \epsilon\left(\frac{\partial \tau}{\partial h}h_1 + \frac{\partial \tau}{\partial g}g_1\right) \bigg|_{h=h_0,g=0}
\end{equation}
hence
\begin{equation}
\begin{gathered}
	\tau_0(z) = \tau(z;0,h_0); \hspace{.25cm}
	\tau_{\epsilon}(z) = \left(\frac{\partial \tau}{\partial h}h_1 + \frac{\partial \tau}{\partial g}g_1\right)\bigg|_{h=h_0,g=0},
\end{gathered}
\end{equation}
and the same convention will be adopted for $\alpha_{10}(z)$ and $\alpha_{1\epsilon}(z)$.

%\begin{equation}
%\begin{gathered}
%	\Bigg[  \frac{asdadasd}{asdasdasd} \\
%	asdasd \Bigg] \\
%	\Bigg\{ adasd \\ \Bigg\}
%\end{gathered}
%\end{equation}

\paragraph{Dispersion relation as a functional: full result.} With the expansions described above, we may obtain a dispersion relation for any prescribed depth-dependence profiles and relation between the interfaces. Because we wish to develop results in the greatest possible generality, we have performed our calculations with all fields $\rho, \vec{v}, \Delta, p$ expressed in terms the depth-dependence profiles, interface relation, and the arbitrary components of the stress tensor. This one-time calculation then provides a highly general stability result in the form of a \textit{functional}, or a ``function of functions", which accepts as its arguments functions rather than numbers and returns a scalar quantity. Hence we report $\sigma = \sigma(\tau,\alpha,D_{ij},\frac{g_1}{h_1},k)$. By the long-wave assumption, we therefore have
\begin{equation}
	\sigma \approx 0 + k\left(\sigma_{10} + \hat{\alpha}\sigma_{11} \right) + k^2\left(\sigma_{20} + \hat{\alpha}\sigma_{21} \right),
\end{equation}
where $\sigma$ is the linear dispersion relation, $k$ is the wavenumber, and $\hat{\alpha}$ is the baseline magnitude of the isotropic swelling mechanism across the film. The analysis detailed in the Appendix leads to
%\begin{equation}
%	\sigma_{10} = -2fAiD_{13}\left( \int_0^{h_0}\int_0^{\hat{z}}(\tau_h + \tau_g\frac{\tilde{g}_1}{\tilde{h}_1})d\tilde{\tilde{z}}d\tilde{z} -\tau(0;0,h_0)\frac{\tilde{g}_1}{\tilde{h}_1}h_0 + \int_0^{h_0}\tau d\tilde{z}  \right),
%\end{equation}
\begin{equation} \label{lsaresult1}
\begin{gathered}
	\sigma_{10} = -\frac{2fAiD_{13}}{\tilde{h}_1}\Bigg[\int_0^{h_0}\int_0^{z_1}\tau_{\epsilon}(z_2)dz_2dz_1 - \tau_0(0)\tilde{g}_1h_0 + \int_0^{h_0}\tau_0(z)dz\Bigg];
\end{gathered}
\end{equation}

%\begin{equation}
%	\begin{gathered}
%		\sigma_{11} = \frac{1}{V}\int_{0}^{h_0}\{
%		-2fAiD_{13} \alpha_1 [\int_0^{\hat{z}}\int_0^{\tilde{z}}(\tau_h + \tau_g \frac{\tilde{g}_1}{\tilde{h}_1})d\tilde{\tilde{z}}d\tilde{z} - \tau(0;0,h_0)\frac{\tilde{g}_1}{\tilde{h}_1}\hat{z}] \\
%		+ \sigma_{10}[\int_0^{\hat{z}}(\frac{\partial \alpha_1}{\partial g}\frac{\tilde{g}_1}{\tilde{h}_1} + \frac{\partial \alpha_1}{\partial h})d\tilde{z} - \alpha_1(0;0,h_0)\frac{\tilde{g}_1}{\tilde{h}_1}]\}d\hat{z},
%	\end{gathered}
%\end{equation}

\begin{equation}
\begin{gathered}
\sigma_{11} = \frac{1}{V\tilde{h}_1}\int_0^{h_0}\Bigg[-2fAiD_{13}\alpha_{10}(z)\Big[\int_0^{z}\int_0^{z_1}\tau_{\epsilon}(z_2)dz_2dz_1 - \tau_0(0)\tilde{g}_1z\Big] \\
+ \sigma_{10}\Big[\int_0^{z}\alpha_{1\epsilon}(z_1)dz_1 - \alpha_{10}(0)\tilde{g}_1\Big]\Bigg]dz;
\end{gathered}
\end{equation}

%\begin{equation}
%	\sigma_{20} = 2fA(D_{11}-D_{33})\int_0^{h_0}\left(\int_{0}^{\hat{z}}\int_0^{\tilde{z}}(\tau_h + \tau_g \frac{\tilde{g_1}}{\tilde{h}_1})d\tilde{\tilde{z}}d\tilde{z} -\hat{z}(\tau(h_0; 0,h_0) + \int_0^{h_0}(\tau_h +\tau_g \frac{\tilde{g}_1}{\tilde{h}_1})d\tilde{\tilde{z}})\right) d\hat{z}, 
%\end{equation}

\begin{equation}
\begin{gathered}
\sigma_{20} = \frac{2fA(D_{11}-D_{33})}{\tilde{h}_1}\int_0^{h_0}\Bigg[\int_0^{z}\int_0^{z_1}\tau_{\epsilon}(z_2)dz_2dz_1 - z\Big[\tau_0(h_0) + \int_0^{h_0}\tau_{\epsilon}(z)dz\Big] \Bigg]dz;
\end{gathered}
\end{equation}
and
%\begin{equation}
%	\begin{gathered}
%		\sigma_{21} = -\int_0^{h_0}\{\frac{\sigma_{10}}{V^2} \int_0^{\hat{z}}[[\int_0^{\tilde{\tilde{z}}}(\frac{\partial\alpha_1}{\partial g}\frac{\tilde{g}_1}{\tilde{h}_1} + \frac{\partial \alpha_1}{\partial h})d\tilde{z} - \alpha_1(0;0,h_0)\frac{\tilde{g}_1}{\tilde{h}_1}][\sigma_{10} + 2fAiD_{13}\int_0^z\tau d\tilde{z}] \\
%		- 2fAiD_{13}\alpha_1(z;0,h_0)[\int_0^z\int_0^{\tilde{z}}(\tau_h + \tau_g\frac{\tilde{g}_1}{\tilde{h}_1})d\tilde{\tilde{z}}d\tilde{z} - \tau(0;0,h_0)\frac{\tilde{g}_1}{\tilde{h}_1}\hat{z}]]d\hat{\hat{z}} \\
%		- \frac{\sigma_{20}}{V}[\int_0^{\hat{z}}(\frac{\partial \alpha_1}{\partial g}\frac{\tilde{g}_1}{\tilde{h}_1} + \frac{\partial \alpha_1}{\partial h})d\tilde{z} - \alpha_1(0;0,h_0)\frac{\tilde{g}_1}{\tilde{h}_1}] \\
%		-[\int_0^{\hat{z}}\int_0^{\tilde{z}}(\frac{\partial \alpha_1}{\partial g}\frac{\tilde{g}_1}{\tilde{h}_1} + \frac{\partial \alpha_1}{\partial h})d\tilde{\tilde{z}}d\tilde{z} + z[\alpha_1(0;0,h_0)\frac{\tilde{g}_1}{\tilde{h}_1} - 2\alpha_1(h_0;0,h_0) - 2\int_0^{h_0}(\frac{\partial \alpha_1}{\partial g}\frac{\tilde{g}_1}{\tilde{h}_1} + \frac{\partial \alpha_1}{\partial h})d\tilde{z}]] \\
%		-\frac{2fA\alpha_1(D_{11}-D_{33})}{V}[\int_0^{\hat{z}}[\int_0^{\tilde{\tilde{z}}}\int_0^{\tilde{\tilde{\tilde{z}}}}(\tau_h + \tau_g \frac{\tilde{g}_1}{\tilde{h}_1})d\tilde{\tilde{z}}d\tilde{z} -z[\tau(h_0;0,h_0) + \int_0^{h_0}(\tau_h + \tau_g\frac{\tilde{g}_1}{\tilde{h}_1})d\tilde{z}]]d\hat{\hat{z}}]\}d\hat{z}.
%	\end{gathered}
%\end{equation}

\begin{equation} \label{lsaresult2}
\begin{gathered}
	\sigma_{21} = -\frac{1}{\tilde{h}_1}\int_0^{h_0}\Bigg[\frac{\sigma_{10}}{V^2}\int_0^{z}\Big[ \Big(\int_0^{z_1}\alpha_{1\epsilon}(z_2)dz_2 - \alpha_{10}(0)\tilde{g}_1\Big)\Big(\sigma_{10} + 2fAiD_{13}\int_0^{z_1}\tau_0(z_2)dz_2\Big) \\
	- 2fAiD_{13}\alpha_{10}(z_1)\Big(\int_0^{z_1}\int_0^{z_2}\tau_{\epsilon}(z_3)dz_{2} - \tau_0(0)\tilde{g}_1z_1\Big)\Big]dz_1 \\
	-
	\frac{\sigma_{20}}{V}\Big[\int_0^z\alpha_{1\epsilon}(z_1)dz_1 - \alpha_{10}(0)\tilde{g}_1\Big] \\
	- \Big[\int_0^z\int_0^{z_1}\alpha_{1\epsilon}(z_2)dz_2dz_1 + z\Big(\alpha_{10}(0)\tilde{g}_1 - 2\alpha_{10}(h_0) - 2\int_0^{h_0}\alpha_{1\epsilon}(z)dz\Big)\Big] \\
	- \frac{2fA\alpha_{10}(z)(D_{11}-D_{33})}{V}\Big[\int_0^z\Big(\int_0^{z_1}\int_0^{z_2}\tau_{1\epsilon}(z_3)dz_3dz_2 - z_1(\tau_0(h_0) + \int_0^{h_0}\tau_{1\epsilon}(z)dz)\Big)dz_1\Big]\Bigg]dz
\end{gathered}
\end{equation}

\paragraph{Dispersion relation as a functional: small cross-terms result.} From previous parameter estimates \cite{madi-thesis-2011,george-etal-JAP-2010,van-dillen-etal-PRB-2005-viscoelastic-model,norris-etal-SREP-2017} and calculations of erosion rates (which are well-established \cite{yamamura-etal-1983-IPP,yamamura-etal-RE-1987,wesch-wendler-book-2016}), we anticipate that the values of $fA$ and $\hat{\alpha}$ should be small, and $V$ large in comparison. This is unsurprising due to the two timescales involved in the problem: $V$ is driven by prompt-regime erosion, while $fA$ and $\hat{\alpha}$ are associated with gradual-regime stress modification. A significant simplification is obtained when we consider that products $fA \hat{\alpha}$ are small. We find

%\begin{equation}
%	\begin{gathered}
%		\sigma_{10} = -2fAiD_{13}\left( \int_0^{h_0}\int_0^{\hat{z}}(\tau_h + \tau_g\frac{\tilde{g}_1}{\tilde{h}_1})d\tilde{\tilde{z}}d\tilde{z} -\tau(0;0,h_0)\frac{\tilde{g}_1}{\tilde{h}_1}h_0 + \int_0^{h_0}\tau d\tilde{z}  \right), \\
%		\sigma_{11} = 0, \\
%		\sigma_{20} = 2fA(D_{11}-D_{33})\int_0^{h_0}\left(\int_{0}^{\hat{z}}\int_0^{\tilde{z}}(\tau_h + \tau_g \frac{\tilde{g_1}}{\tilde{h}_1})d\tilde{\tilde{z}}d\tilde{z} -\hat{z}(\tau(h_0; 0,h_0) + \int_0^{h_0}(\tau_h +\tau_g \frac{\tilde{g}_1}{\tilde{h}_1})d\tilde{\tilde{z}})\right) d\hat{z}, 			
%	\end{gathered}	
%\end{equation}

\begin{equation}
	\begin{gathered}
		\sigma_{10} = -\frac{2fAiD_{13}}{\tilde{h}_1}\Bigg[\int_0^{h_0}\int_0^{z_1}\tau_{\epsilon}(z_2)dz_2dz_1 - \tau_0(0)\tilde{g}_1h_0 + \int_0^{h_0}\tau_0(z)dz\Bigg];
	\end{gathered}
\end{equation}

\begin{equation}
	\sigma_{11} = 0;
\end{equation}

\begin{equation}
	\sigma_{20} = \frac{2fA(D_{11}-D_{33})}{\tilde{h}_1}\int_0^{h_0}\Bigg[\int_0^{z}\int_0^{z_1}\tau_{\epsilon}(z_2)dz_2dz_1 - z\Big[\tau_0(h_0) + \int_0^{h_0}\tau_{\epsilon}(z)dz\Big]\Bigg]dz
\end{equation}
and
%\begin{equation}
%	\begin{gathered}
%		\sigma_{21} = \int_0^{h_0}\left(
%		\left(\int_0^{\hat{z}}\int_0^{\tilde{z}}(\frac{\partial \alpha_1}{\partial g}\frac{\tilde{g}_1}{\tilde{h}_1} + \frac{\partial \alpha_1}{\partial h})d\tilde{\tilde{z}}d\tilde{z} + z[\alpha_1(0;0,h_0)\frac{\tilde{g}_1}{\tilde{h}_1} - 2\alpha_1(h_0;0,h_0) - 2\int_0^{h_0}(\frac{\partial \alpha_1}{\partial g}\frac{\tilde{g}_1}{\tilde{h}_1} + \frac{\partial \alpha_1}{\partial h})d\tilde{z}]\right)
%		\right)dz,
%	\end{gathered}
%\end{equation}

\begin{equation}
\begin{gathered}
	\sigma_{21} = \frac{1}{\tilde{h}_1}\int_0^{h_0}\Bigg[\int_0^z\int_0^{z_1}\alpha_{1\epsilon}(z_2)dz_2dz_1 +z\Big[\alpha_{10}(0)\tilde{g}_1 - 2\alpha_{10}(h_0) - 2\int_0^{h_0}\alpha_{1\epsilon}(z)dz\Big]     \Bigg],
\end{gathered}
\end{equation}
which we will use throughout the rest of the present work.

\paragraph{The interface relation.} 
Using the expressions above, we may study the value $\theta_c$ for which $\sigma$ transitions from negative (i.e., the surface is stable to perturbations) to positive (i.e., the surface is unstable to perturbations) for given depth-dependence profiles $\tau(z;g,h), \alpha(z;g,h)$ and \textit{interface relation} $\frac{\tilde{g}_1}{\tilde{h}_1}$, which together can be used to study many physical systems in great generality. We take a moment to characterize this last quantity, which has seen only little explicit treatment to date \cite{Swenson_2018,evans-norris-JPCM-2022}.

As a thin-film hydrodynamic stability problem, any analysis requires some notion of the lower interface. Elsewhere, various assumptions about the relationship between the upper interface and lower interface have been explored, including a flat lower interface \cite{castro-cuerno-ASS-2012}, a lower interface that is a vertical translation of the upper interface \cite{norris-PRB-2012-linear-viscous,norris-PRB-2012-viscoelastic-normal}, and a lower interface that shifts horizontally as $\sin(\theta)$ and thins vertically as $\cos(\theta)$ \cite{moreno-barrado-etal-PRB-2015,Swenson_2018,evans-norris-JPCM-2022}. Physically, these interface relations are rooted in the understanding that the ion-beam permits a certain amorphization thickness and, as material is sputtered away, this exposes more material to amorphization along the ion-beam direction. This dependence naturally appears in the linearization of the interfaces and is a fundamental quantity in the physical description of the system. We have therefore left the expressions for the perturbations to the interfaces, $\tilde{g}_1$ and $\tilde{h}_1$, arbitrary in the calculations in the Appendix. Notice that these terms only ever occur in the ratio $\frac{\tilde{g}_1}{\tilde{h}_1}$. This quantity, then, captures the way that perturbations to the upper interface influence the lower interface. Suppose that we had the expressions
\begin{equation}
	\begin{gathered}
		\tilde{h}_1 = \exp(\sigma t + ikx) \\
		\tilde{g}_1 = \exp(\sigma t + ik(x-x_0))
	\end{gathered}	
\end{equation}
in the original linearization. Then we have 
\begin{equation}
	\frac{\tilde{g}_1}{\tilde{h}_1} = \exp(-ikx_0),
\end{equation}
which appears as a parameter in the equations of linearization. After evaluating (equation ref) with the ratio replaced by the right-hand side of the above, we may expand the above in $k\approx 0$, collect terms, and finally obtain a dispersion relation which retains the desired interface relation. Since $x_0$ is the horizontal shift, taking $x_0=0, h_0=\text{constant}$ imparts the vertical-translation interface relation upon the system. Likewise, $x_0(\theta)=h_0(0)\sin(\theta), h_0(\theta)=h_0(0)\cos(\theta)$ imparts the diagonal-translation, and $\tilde{g}_1 = 0$ (hence the elimination of all $\frac{\tilde{g}_1}{\tilde{h}_1}$ terms) imparts the flat lower interface assumption.

\paragraph{Legendre polynomials in downbeam-crossbeam coordinates.}

\begin{figure}[h!]\label{ionschematic}
	\centering 
	\includegraphics[totalheight=5cm]{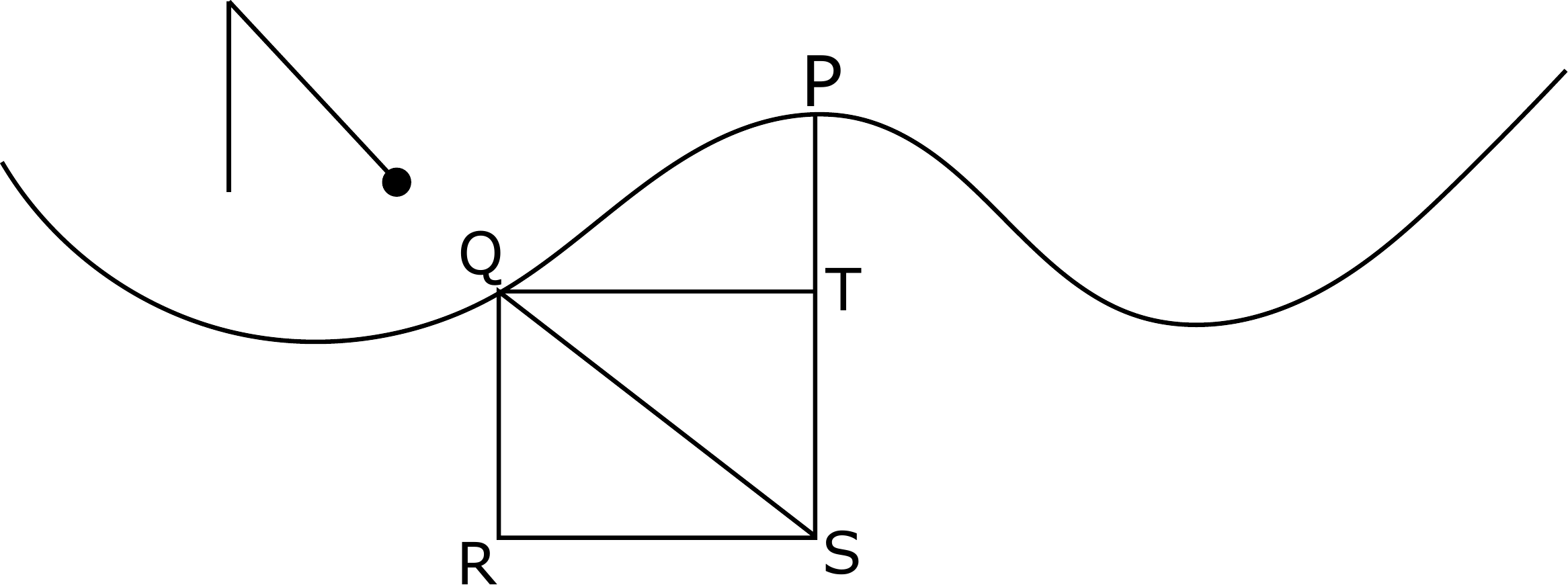}
	\caption{Schematic depicting the construction of depth-dependence at a point $S$ in the film depth in Cartesian $x,z$ coordinates by mapping the influence of an ion that enters the surface at the point $Q$ to the point $S$. In the limit of small slopes, this geometric argument leads to analytically-tractable forms.}
\end{figure}

Choose an arbitrary patch of surface (X,h(X)) on a surface (x,h(x)), identified with point $P$ in Figure 2 and examine some distance $z$ below its surface, $(X,h(X)-z)$, identified with $S$. Given an irradiation angle $\theta$ (from the left) and under the assumption of no deflection of that angle within the film, the point $(X,h(X)-z)$ (``S") is associated with precisely one location on a nearby patch of surface from which the deposition occurring at $(X,h(X)-z)$ originates. Denote this associated patch of surface as the point $(x_1, h(x_1))$, identified with the point Q, the ``entry point" at which an ion passes through the surface before initiating a collision cascade. Construct a triangle using the points $(X,h(X)-z)$ (S), $(x_1, h(x_1))$ (Q) and an additional point determined by the intersection of the vertical line running through $x_1$ and the horizontal line $h(X)-z$ (R), i.e., the triangle $\triangle RQS$, whose interior angle is the same as that of the angle of the incoming ion to the laboratory frame's $z$ axis. The lengths of this triangle can be easily determined exactly via right-angle trigonometry. With the Taylor expansion of the surface about $x=X$, we may approximate near $x=x_1$. and with $x_1 = X - x_2$ by definition, we have
\begin{equation}
	h(x_1) = h(X-x_2) = h(X)-x_2\frac{\partial h}{\partial x} + ...,
\end{equation}
which implies that line segment $QR$ is approximately of length $(h(X)-z) - x_2\frac{\partial h}{\partial x}$. We now seek the horizontal distance from the entry point $(x_1,h(x_1))$ to the point of interest, $(X, h(X)-z)$, the line segment $RS$. This horizontal shift, $x_2$, whose length is identified with the line segment $RS$ is precisely the length of the right triangle opposite the interior angle $\theta$ and is immediately given by 
\begin{equation}
	x_2 = \frac{\tan(\theta)(h(X)-z)}{1 + \tan(\theta)\frac{\partial h}{\partial x}|_{X}}
\end{equation}
Finally, the hypotenuse (the line segment QS), equivalent to the downbeam distance, which we will denote $\tilde{z}$, has the form
\begin{equation}
	\tilde{z} = \sec(\theta)\left(h(X)-z - x_2\frac{\partial h}{\partial x}\right)\Bigg|_{X}
\end{equation}
which can be used to compute the influence of an arbitrary Legendre polynomial defined in downbeam coordinates. That is, we may express the depth-dependence profile along the line containing $P$ and $S$ in terms of a down-beam stress profile originally defined along the line containing $Q$ and $S$. For a given triplet of surface $h(X)$, ion-angle $\theta$ and point in the bulk $S$, there is \textit{exactly one} patch of surface (the point $Q$) which determines the influence of the ion beam at the point $S$, hence the transformation is invertible and all depth-dependence throughout the entire film is uniquely assigned at each arbitrary point $S$ of the bulk. From the above argument, we find that in the limit of small slopes, the entire transformation is expressible in a manner suited to the present linear stability analysis. This construction is applicable to any arbitrary $(X,h(X))$; we now drop the dummy variable and consider
\begin{equation}
	\tau(\hat{z}) = \sum_{k=0}^N \beta_k P_k(\hat{z}),
\end{equation}
where $P_k(\hat{z})$ are the typical Legendre polynomials defined over $[-1,1]$. We may then transform to an arbitrary setting via
\begin{equation}
	\hat{z} \to \frac{2\tilde{z} - (\hat{g}+\hat{h})}{\hat{h}-\hat{g}},
\end{equation}
so that $\hat{g}$ is mapped to $-1$, and $\hat{h}$ is mapped to $1$. We now set $\hat{g} = 0$ and $\hat{h} = h_0'$, the fixed downbeam distance as seen from downbeam coordinates. Hence
\begin{equation}
\begin{gathered}
		\tau(z) = \sum_{k=0}^N \beta_k P_k(\frac{2\tilde{z}-h_0'}{h_0'}),
\end{gathered}
\end{equation}
where
\begin{equation}
\begin{gathered}
		\tilde{z} = \frac{(h-z)}{\cos(\theta) + h_x\sin(\theta)}
	\end{gathered}
\end{equation}
expresses the influence of an arbitrary Legendre polynomial, originally defined in downbeam coordinates, on an arbitrary location (in terms of the Cartesian $z$ axis, as opposed to the downbeam $\tilde{z}$) while the small-slopes approximation holds. For the reader's convenience, we note that the typical $n^{th}$ Legendre polynomial $P_n(x)$ comprises an orthogonal family on the interval [-1,1], and is easily computed by the Rodrigues formula,
\begin{equation}
	P_n(x) = \frac{1}{2^n n!}\frac{d^n}{dx^n}\left[\left(x^2 - 1\right)^n\right].
\end{equation}
The first three such polynomials are therefore
\begin{equation}
\begin{gathered}
 P_0(x) = 1, \\
 P_1(x) = x, \\
 P_2(x) = \frac{1}{2}\left(3x^2-1\right).
\end{gathered}
\end{equation}

\paragraph{Conditions for positivity of weighted sums up to $P_2$.} We want to ensure that the Legendre polynomials are positive on the domain of concern. Otherwise, we would be computing dispersion relations associated with ``negative APF" or ``negative swelling", which are nonphysical. One set of parameter values is obtained by simply requiring that \textit{all} roots of the polynomial are complex. Hence, given the weighted sum of the first three terms in the sequence of Legendre polynomials,
\begin{equation}
	1 + \beta_1 x + \frac{\beta_2}{2}(3x^2-1),
\end{equation}
we require both
\begin{equation}
\begin{gathered}
	-\sqrt{3} < \beta_1 < \sqrt{3} \\
	\text{and} \\
	1 - \frac{\sqrt{3-\beta_1^2}}{\sqrt{3}} < \beta_2 < 1 + \frac{\sqrt{3-\beta_1^2}}{\sqrt{3}},
\end{gathered}
\end{equation}
which forces the discriminant to be negative and real, leading to complex roots only. This is an ellipse in $\beta_1, \beta_2$ parameter space and is associated with $\beta_2 \geq 0$; this says that the influence of the beam is at a minimum in the center of the film, which seems unlikely based on our suspicions about the underlying physics. When $\beta_2 \neq 0$, the singleton $x= -\frac{\beta_1}{3\beta_2}$ occurs where the discriminant is zero, which occurs precisely for
\begin{equation}
	\beta_1^2 + 3(\beta_2-1)^2 = 3,
\end{equation}
and we then have the double-root $x=-\frac{\beta_1}{3\beta_2}$, which can be forced outside of the interval $[-1,1]$ by either of $\beta_2 < \frac{-\beta_1}{3}$ or $\beta_2 > \frac{\beta_1}{3}$. Hence the intersections of the closed set characterized by the ellipse and the open set characterized by the regions above the lines are regions where the values of $(\beta_1,\beta_2)$ guarantee that the double-root occurs outside of $[-1,1]$. If $\beta_2 = 0$, we simply have $\beta_1 > -1$ or $\beta_1 < 1.$ The final set is obtained by taking $\beta_2 \neq 0$, allowing two real roots, such that the discriminant is positive, but requiring that they \textit{both} occur outside of the interval [-1,1]. A straightforward idea is to obtain roots for $x$ from the quadratic formula, impose $|x| > 1$, and work through the various cases involving $\beta_1,\beta_2$. The calculation is tedious due to the multiple cases involved, but we ultimately obtain our final solution set,
\begin{equation}
\begin{gathered}
	\textbf{S} = \{(\beta_1,\beta_2) \in \textbf{R}^2 : \beta_1^2 + 3\left(\beta_2 - 1\right)^2 < 3\} \bigcup \{(\beta_1,\beta_2) \in \textbf{R}^2 : |\beta_1|-1 < \beta_2 < 1 - \frac{\sqrt{3-\beta_1^2}}{\sqrt{3}}\}
\end{gathered}
\end{equation}
for which $\sum_{k=0}^2 \beta_k P_k(x)$ is positive over $x \in [-1,1]$. In the $(\beta_1,\beta_2)$ plane, this is an ``ice-cream cone shape". We note that a full characterization of the coefficients $\{\beta_1, ..., \beta_N\}$ such that $\sum_{k=0}^N\beta_kP_k(x) \geq 0$ for $x \in [-1,1]$ is highly nontrivial and a solution attempt requires techniques from algebraic geometry, being related to the idea of \textit{semialgebraic sets}. It appears that no general result yet exists for this problem. Because we have invoked Legendre polynomials only as a stand-in for a more detailed depth-dependence profile, we do not pursue it here.

\paragraph{A note on flux dilution and ``zero-width" ion tracks.} 
It is illustrative to note that the assumption of stress modification that varies along the downbeam axis is essentially the limiting case of stress modification due to an intensity field generated by a Gaussian ellipsoid (as in the analysis of \cite{sigmund-PR-1969,sigmund-JMS-1973,bradley-harper-JVST-1988,}) as the cross-beam width approaches zero. We may consider arbitrary function $f$ of downbeam coordinate $\hat{z}$ and its product with a Gaussian in crossbeam coordinate $\hat{x}$ such that
\begin{equation}
	\begin{gathered}
		\hat{z} = (x-X)\sin(\theta) - (z-h(X))\cos(\theta) \\
		\hat{x} = (x-X)\cos(\theta) + (z-h(X))\sin(\theta),
	\end{gathered}
\end{equation}
where $z=h(X)$ is the surface (in Cartesian), X is the location of the incoming ion (in Cartesian). We will use $(x,z)$ to specify the point at which we want to compute instantaneous energy deposition, and integration across all $X$ locations from which ions can enter will give average power deposition at an arbitrary point. We will allow $\beta$ to denote the standard deviation of a Gaussian describing the crossbeam component of this power deposition function, assumed to be separable in downbeam-crossbeam coordinates. With the inclusion of geometric flux dilution, we have
\begin{equation}
	\begin{gathered}
		P(\hat{x},\hat{z};X,h(X)) = \frac{1}{\sqrt{2\pi}\beta}\int_{-\infty}^{\infty} \left(\cos(\theta) + h_X\sin(\theta)\right) f(\hat{z})\exp\left(-\frac{[(x-X)\cos(\theta) + (z-h(X))\sin(\theta)]^2}{2\beta^2}\right)dX.
	\end{gathered}
\end{equation}
We notice that in the limit as $\beta \to 0$,
\begin{equation}
	\frac{1}{\sqrt{2\pi}\beta}\exp\left(-\frac{[(x-X)\cos(\theta) + (z-h(X))\sin(\theta)]^2}{2\beta^2}\right) \to \delta\left((x-X)\cos(\theta) + (z-h(X))\sin(\theta) \right),
\end{equation}
because
\begin{equation}
	\begin{gathered}
		\delta(x) = \lim_{c \to 0} \frac{1}{c\sqrt{\pi}}e^{-x^2/c^2} 
	\end{gathered}
\end{equation}
by (one) definition, and
\begin{equation}
	\delta(\frac{x}{\sqrt{2}}) = \lim_{c \to 0} \frac{1}{c\sqrt{\pi}}e^{-x^2/(2c^2)} = \sqrt{2}\delta(x)
\end{equation}
by the scaling property. We make the substitution $w = -[(x-X)\cos(\theta) + (z-h(X))\sin(\theta)]$ so that
\begin{equation}
	P(\hat{x},\hat{z};X,h(X)) = \int_{-\infty}^{\infty}f(\hat{z})\delta\left(w\right) dw,
\end{equation}
and $w=0$ (i.e., $\hat{x}=0$) implies $f(\hat{z}) = (h-z)\sec(\theta)$ after simplification, so
\begin{equation}
	P(\hat{x},\hat{z};X,h(X)) = f\left(\hat{z}\bigg|_{w=0}\right) = f\left((h(X_0)-z)\sec(\theta)\right),
\end{equation}
where $X_0$ is the specific value of $X$ that coincides with the ion depositing energy along $w = \hat{x} = 0$.
and this can be linearized to determine $P(x,z;X,h(X))$, i.e., conversion into Cartesian from downbeam-crossbeam. Flux dilution has been canceled out exactly by the vanishing of the $\hat{x}$ coordinate. We will still need to approximate $h(X_0)$ near $x$ in order to compute values at $(x,z)$ for the purposes of linear stability analysis, and such calculations will be inserted into the pre-computed functional form of the dispersion relation described in Section 3.1. This observation will nonetheless simplify our work in the idealized case of stress modification along vanishingly-thin ion tracks.

\subsection{Parameter estimation: 250eV Ar$^+ \to$ Si}
\paragraph{Notation for parameters.}
In the present work, we have used $fA$ as a coefficient of the APF-term and $\hat{\alpha}$, borrowing from the notation of \cite{van-dillen-etal-PRB-2005-viscoelastic-model,otani-etal-JAP-2006,madi-thesis-2011,norris-PRB-2012-linear-viscous}. As a coefficient of the IIS-term, we have used $\hat{\alpha}$, notation similar to that of \cite{Swenson_2018,evans-norris-JPCM-2022}. Both quantities have natural units of $\frac{1}{\text{s}}$, being, fundamentally, \textit{rates}. From the work of \cite{ishii-etal-JMR-2014,norris-etal-SREP-2017,NorrisAziz_predictivemodel}, it is expected that these rates should vary according to flux; we may then expect that there should exist an $A_I$, representing the isotropic component of stress due to the ion beam, in the same way that there is an $A_D$ that represents the deviatoric component of stress. Hence we will consider
\begin{equation}
	fA \to fA_{D}, \hspace{.25cm} \hat{\alpha} \to fA_{I},
\end{equation}
with subscripts denoting the deviatoric and isotropic parts of deformation respectively. This makes the assumption that both terms scale with flux. While this seems probable, too little is known about the underlying physics at the present time. We therefore adopt this notation on a tentative basis.

\paragraph{Estimation of viscosity $\eta$.}
From our analysis, we consider the top-left component of the steady-state stress tensor, which will permit comparison with experimental data. We find
\begin{equation}
	\begin{gathered}
		T_{0}^{11} = -p_{0} -2fA\eta\tau(z;0,h_0)D_{11} = \\
		-(p_{00} +\hat{\alpha} p_{01}) - 2fA\eta\tau(z;0,h_0)D_{11}.
	\end{gathered}
\end{equation}
With $p_{00} = -2fA\eta\tau(z;0,h_0)D_{33}$ and $p_{01} = 2\eta \alpha_1$, we have
\begin{equation}
	\begin{gathered}
		T_{0}^{11} = -2fA\eta\tau(z;0,h_0)(D_{11}-D_{33}) - 2\hat{\alpha}\eta \alpha_1(z;0,h_0) \\ = \\ -6fA\eta\cos(2\theta)\tau(z;0,h_0) - 2\hat{\alpha}\eta \alpha_1(z;0,h_0).
	\end{gathered}
\end{equation}
However, because the steady-state in-plane stress includes the forms $fA\eta$ and $\alpha \eta$, we also require an estimate of $\eta$ for the 250eV Ar$^+ \to$ Si system. Toward this end, we note that previous work described in \cite{madi-thesis-2011} has estimated $\eta$ for the 250eV Ar$^+ \to$ Si as $\eta \approx 6.2\times10^{-1} \text{GPa} \cdot \text{s}$ for a flux $f = 1.2\times 10^{1} \frac{\text{ions}}{\text{nm}^2\cdot \text{s}}$. However, Madi \cite{madi-thesis-2011} compares this estimate for viscosity with his experimental results and finds that it suggests a \textit{vastly} different relaxation time than experimentally observed. This is well-aligned with the discussion in \cite{perkinsonthesis2017} which finds relaxation times for experimental data of 250eV Ar$^+$ on Si consistent, within an order of magnitude, with the estimate of $\eta \approx 1.5\times 10^2 \text{ Gpa} \cdot \text{s}$ from \cite{norris-etal-SREP-2017}, despite the estimate of the latter having been fit for 1keV Ar$^+$ on Si. This may suggest that $\eta$ does not vary strongly with energy for the same ion and target species. We therefore tentatively adopt the estimate of \cite{norris-etal-SREP-2017}, an established value for which relaxation times are roughly in agreement with that of the present experimental system.

\paragraph{Estimation of ion-induced stress components.} 
We may then use the theoretical steady-state mean in-plane stress
\begin{equation}
 <T_{0}^{11}> = \frac{1}{h_0}\int_0^{h_0}T_{0}^{11}d\tilde{z}
 \end{equation}
for comparison with experimental data \cite{perkinsonthesis2017} of wafer curvature measurements. We here note that we have taken down-beam depth-dependence profiles which, in the steady state, reduce to the form
\begin{equation}
	\alpha_1(z;0,h_0) = \tau(z;0,h_0) = 1 + \beta_1 P_1\left(\frac{h_0-2z}{h_0}\right) + \beta_2P_2\left(\frac{h_0-2z}{h_0}\right),
\end{equation}
and we have
\begin{equation}
	<\alpha_1(z;0,h_0)> = <\tau(z;0,h_0)> = 1
\end{equation}
for all $\beta_1,\beta_2$. We therefore seek to fit
\begin{equation} \label{txxcomponent}
	\begin{gathered}
		<T_{0}^{11}> = -2fA\eta(D_{11}-D_{33}) - 2\hat{\alpha}\eta  \\ = \\ -6fA\eta\cos(2\theta) - 2\hat{\alpha}\eta
	\end{gathered}
\end{equation}
to the experimental data of Perkinson.

\begin{figure}[h!]
	\centering
	\includegraphics[totalheight=8cm]{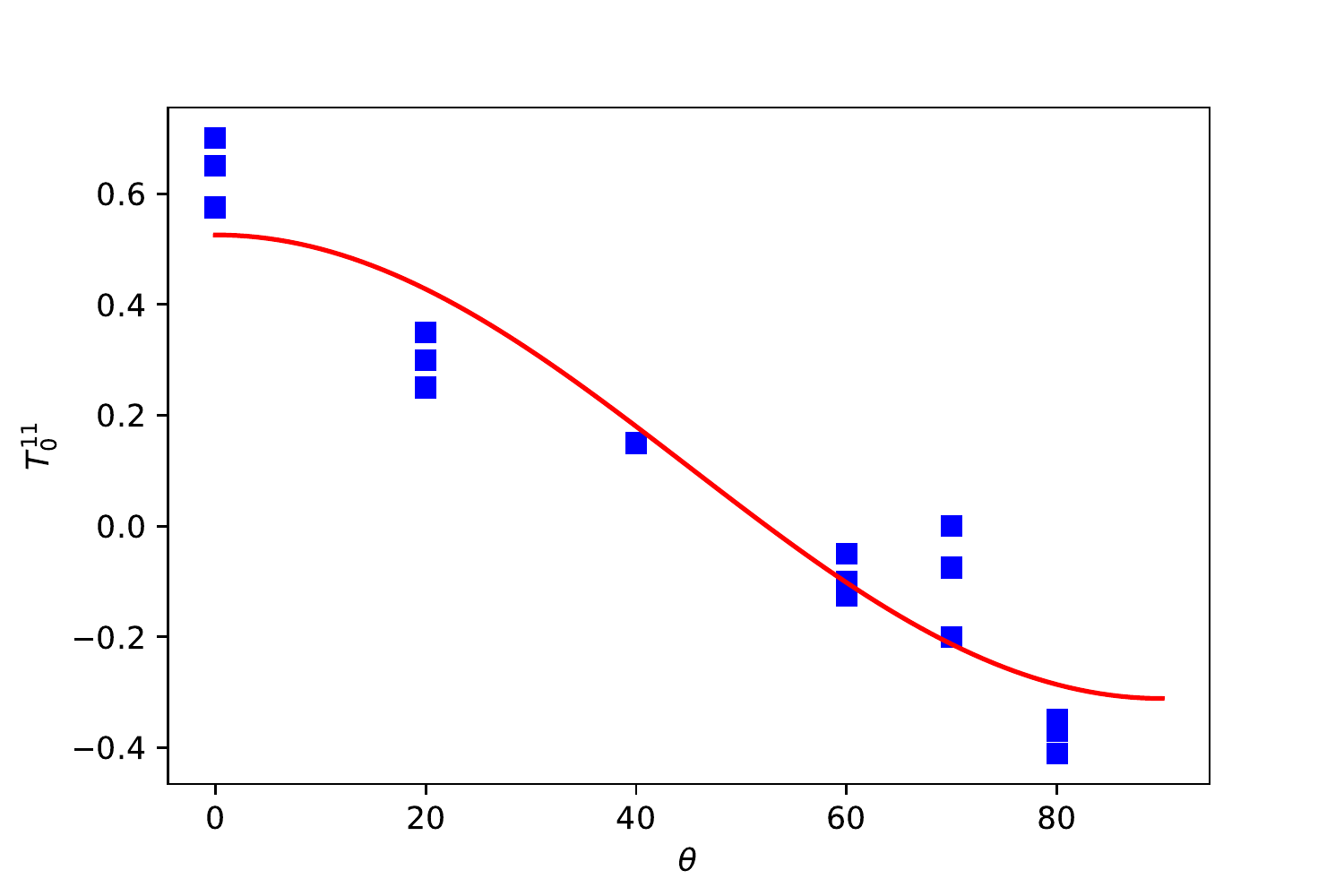}
	\caption{Fitting the form of Equation (\ref{txxcomponent}) to the data of \cite{perkinsonthesis2017}, which was collected using a machine of design and construction original to that work. \textbf{Blue:} experimental data set. \textbf{Red:} the theoretical magnitude of $T_{0}^{11}$, the steady-state in-plane component of stress, when fit to the data. Because $<T_0^{11}>$ has only two free parameters while the data set consists of three observations at each of six different incidence angles, we may perform parameter-fitting from the data.}
\end{figure}

Other attempts at direct measurement of $A_D$ have been made. In \cite{madi-thesis-2011}, it was estimated that $A_D \approx 1.15\times 10^{-2} \frac{\text{nm}^2}{\text{ion}}$. From George \cite{george-etal-JAP-2010}, we also have an estimate for 3keV Ar$^+ \to$ Si, $A_D \approx 5\times 10^{-3} \frac{\text{nm}^2}{\text{ion}}$. From \cite{norris-etal-SREP-2017} for 1keV Ar$^+ \to$ Si, the parameter group $fA \approx 3\times 10^{-4} \frac{1}{\text{s}}$ at a flux of $f = 2\times 10^{-2} \frac{\text{ions}}{\text{nm}^2\cdot \text{s}}$, implying $A_D \approx 1.5\times 10^{-2} \frac{\text{nm}^2}{\text{ion}}$. All of these estimates are within roughly an order of magnitude, again suggesting small variation for low energies. It is worth pointing out, however, that these estimates, except for that of \cite{norris-etal-SREP-2017, perkinsonthesis2017}, occurred via normal-incidence irradiation; because there was no consideration of isotropic swelling effects nor angle-dependence, all compressive stress at normal incidence was assumed to originate from anisotropic plastic flow. The estimate from \cite{norris-etal-SREP-2017}, on the other hand, was deduced from parameter-fitting to angle-dependent data, and is interpreted as an order-of-magnitude estimate. We again note that the model of \cite{norris-etal-SREP-2017} did not include swelling, took the interface relation to be pure vertical displacement for the purposes of the linear dispersion used to describe APF, and directly added the erosive dispersion relation by way of the results of \cite{bradley-PRB-2011b}, which may further complicate these estimates.

We minimize the $L^2$ error between theoretical steady-state in-plane stress function and angle-dependent stress measurements from Perkinson's data \cite{perkinsonthesis2017} by toggling $fA\eta$ and $\hat{\alpha}\eta$ and tracking propagation of error using the \textit{uncertainties} and $scipy.optimize.curve\_fit$ Python packages. This leads to the estimates
\begin{equation}
	\begin{gathered}
		fA\eta \approx 0.0698 \pm 0.0059 \text{ GPa}; \\
		\hat{\alpha}\eta \approx 0.05346 \pm 0.0132. \text{ GPa}.
	\end{gathered}
\end{equation}
Dividing through by the flux $f = 1.2\times 10^{1} \frac{\text{ions}}{\text{nm}^2\cdot \text{s}}$ (which was the same as that of \cite{madi-thesis-2011}, hence a valid point of comparison) and the above value $\eta \approx 1.5\times 10^2 \text{ GPa} \cdot \text{s}$ from \cite{norris-etal-SREP-2017}, we estimate the parameter group
\begin{equation}
	fA \approx 4.653 \times 10^{-4} \frac{1}{\text{s}},
\end{equation}
as compared with $fA \approx 3\times10^{-3} \frac{1}{\text{s}}$ for 1keV Ar+ on Si in \cite{norris-etal-SREP-2017}. We also obtain
\begin{equation}
	\begin{gathered}
		A_D \approx 3.87\times 10^{-5} \frac{\text{nm}^2}{\text{ion}} \\ 
		A_I \approx 2.97\times 10^{-5} \frac{\text{nm}^2}{\text{ion}}.
	\end{gathered}
\end{equation}
This may be interpreted as a reasonable, order-of-magnitude level of agreement between the present estimates for $A_D$ and those from the experiments of \cite{madi-thesis-2011} and \cite{norris-etal-SREP-2017}, although perhaps lower than expected. On the other hand, we expect that this should be an underestimate of these values, given that we have used a value of $\eta$ associated with higher-energy irradiation. We may therefore have some confidence that this estimate of $A_D$ is reasonable for the present experimental system. 

We note, however, that this estimate is fundamentally reliant on our assumptions about depth-dependence: because we have assumed that both APF and IIS both originate with identical distributions along the ion track, these estimates could be significantly wrong if, for reasons discussed previously, the underlying mechanisms driving each of these two phenomenological effects occur at different magnitudes along the downbeam coordinate. As previously discussed, if it turns out that APF is due to electronic stopping, its intensity should be a decreasing function of downbeam distance, while, if IIS were a result of nuclear stopping power or ion-implantation, we would expect its intensity field in downbeam-crossbeam coordinates to closely resemble that generated by a Gaussian in the downbeam coordinate. This would imply a peak deeper in the film, occurring well-after electronic stopping has become negligible (indeed, it is precisely when enough energy has been lost due to electronic stopping that nuclear stopping becomes dominant within a single cascade event \cite{liedke-thesis-2011}). The difference in the physical locations of these stress-modifying effects would therefore affect the calculation of the above constants. With this caveat in mind, we consider this idealization as a staging point for our current approach.

%\paragraph{Note on Stoney's equation, flux dilution and film thickness.}
%These estimates have been obtained using Stoney's equation,
%\begin{equation}
%	\sigma_{avg} = \frac{\Delta K M_s h_s^2}{6h_f},
%\end{equation}
%where $\sigma_{avg}$ is the averaged in-plane stress within the film, $\Delta K$ is the change in curvature, $M_s$ is the biaxial modulus of the material, $h_s$ is the thickness of the substrate, and $h_f$ is the thickness of the amorphous layer. We note that in the original stress measurements of Perkinson, the film thickness was assumed to be 3nm throughout all irradiation angles; while it is known that the film thickness varies with irradiation angle, we have retained their assumption here. Our steady-state stress tensor includes the influence of flux dilution via a factor of $\cos(\theta)$; with the assumption that the biaxial modulus of the film also scales with the ion flux, hence by $\cos(\theta)$, both factors cancel from the stress measurement. 

\section{Results}
\paragraph{Summary.}After having computed the leading-order real term of the long-wave linear dispersion relation, we determine the beam-angle $\theta_c$ for which the real part changes sign from negative to positive, hence topographical perturbations destabilize the free interface and we may expect pattern formation. We consider the specialization of the dispersion relations stated in Equations (\ref{lsaresult1})-(\ref{lsaresult2}) to each of two cases that have been previously studied. 

Primarily, we consider the case of diagonally-translated interfaces with depth-dependence induced by Legendre polynomials about the fixed downbeam length $h_0'$ that separates the interfaces. For the sake of comparison, we consider vertically-translated interfaces with stress profiles determined about the vertical rather than the downbeam direction. We note that in the case of vertical stress modification at all angles of incidence, there is no meaningful way to establish a connection between flux dilution, beam direction, and the ion-induced stress profile. This is, of course, nonphysical, as we expect stress to be generated at least \textit{approximately} along ion tracks; nonetheless, we consider such a model only because it has been used elsewhere and may be considered the ``opposite extreme" of the diagonal translation case.

\paragraph{Stress modification along downbeam ion tracks.} Even if phenomenological, any analysis of ion-induced stress with depth-dependence about the film requires at least \textit{some} notion of \textit{where} the stress is generated. We argue that the most natural hypothesis is that the stress is generated in the downbeam direction, as this is the direction along which the collision cascade is initiated, and along which the cylindrical, melted ion-track occurs in the electronic stopping regime. It is also natural to incorporate the effect of flux dilution. We have therefore used
\begin{equation}
	\alpha_1(z;g,h,h_x) = \tau(z;g,h,h_x) =  1 + \beta_1 P_1(\frac{2\tilde{z}-h_0'}{h_0'}) + \beta_2P_2(\frac{2\tilde{z}-h_0'}{h_0'}),
\end{equation}
where
\begin{equation}
	\begin{gathered}
		\tilde{z} = \frac{(h-z)}{\cos(\theta) + h_x\sin(\theta)}
	\end{gathered}
\end{equation}
when calculating the following results with stress generated about the ion tracks, a choice which represents the various components of our analysis previously discussed. We remind the reader that $\theta$ is the nominal beam angle in laboratory coordinates, $\tilde{z}$ computes the downbeam distance from a nearby patch of surface to the bulk beneath a given location (so that downbeam stress modification is precise). $P_1, P_2$ are Legendre polynomials, $h_0'$ is a fixed downbeam depth, and $\beta_1,\beta_2$ are free constants which may be used to modify the influence of downbeam stress modification exactly along the downbeam direction. We take the pair of interfacial shifts defined via
\begin{equation}
\begin{gathered}
	h_0(\theta) = h_0'\cos(\theta); \hspace{.25cm} x_0(\theta) = h_0'\sin(\theta); \\ \to 
	\frac{\tilde{g}_1}{\tilde{h}_1} = \exp(-ikx_0(\theta)),
\end{gathered}
\end{equation}
which impart upon the analysis that the film thins as $\cos(\theta)$ with beam angle, and the lower interface is a copy of the upper interface phase-shifted by $x_0(\theta)$. This assumption has been made elsewhere in \cite{moreno-barrado-etal-PRB-2015,Swenson_2018,evans-norris-JPCM-2022}. We obtain the dispersion relation
\begin{equation}
	\sigma = -\frac{1}{6}\left(4\beta_1 - 3(2+\beta_2)\right)\left(3fA(\cos(2\theta) - \sin(2\theta)) + \frac{\hat{\alpha}}{2}\right)\left( kh_0'\cos(\theta)\right)^2 + O((kh_0'\cos(\theta))^4)
\end{equation}
Intriguingly, all depth-dependence factors out cleanly, and the only quantity capable of shifting the selected bifurcation angle is the ratio of $\hat{\alpha}$ to $fA$. We compute the critical angle $\theta_c$,
\begin{equation}
	\theta_c = \frac{1}{2}\arccos\left(\frac{1}{2}\left[1 - \frac{\hat{\alpha}}{6fA} \right]    \right).
\end{equation}

\paragraph{Vertically-displaced interfaces; stress modification vertical.} 
Although nonphysical, we will briefly consider the case that the interfaces are vertically translated and stress-modification, rather than developing along the ion track, as would be natural, instead develops along the vertical axis at all times. Because we no longer have any sense of the beam orientation's relevance to the bulk, the calculation is much simpler, and there is no clear way to incorporate flux dilution. Here, we simply neglect it. We may therefore consider
\begin{equation}
	\begin{gathered}
		\tau(z) = 1 + \sum_{k=1}^N \beta_k P_k(\frac{2z-(g+h)}{(g-h)})
	\end{gathered}
\end{equation}
directly, as downbeam coordinate $\tilde{z}$ is effectively ignored in the ``vertical displacements" case; we instead suppose that the ion-induced stress modification occurs along the laboratory $z$ axis. It is clear that $z=h$ maps to the $\hat{z}=-1$ in the Legendre polynomial's native domain, and $z=g = h-h_0$ maps to $\hat{z} = 1$ in the Legendre polynomial's native domain. This preserves the mapping of the Legendre polynomials from the top of the film to the bottom, analogous to the previous calculations where we considered the downbeam direction as originating somewhere on the upper surface. We then take
\begin{equation}
	\begin{gathered}
		h_0(\theta) = h_0, \text{ fixed}; \hspace{.25cm}
		x_0(\theta) = 0; \\ 
		\to \frac{\tilde{g}_1}{\tilde{h}_1} = 1
	\end{gathered}
\end{equation}
and we compute the dispersion relation
\begin{equation}
	\frac{\text{Re}(\sigma)}{(kh_0)^2} = -(3+\beta_1)\left(fA\cos(2\theta) + \frac{\hat{\alpha}}{6}  \right) + (kh_0)^2,
\end{equation}
and it is clear that the depth-dependence factors out completely, with $\theta_c$ determined fully by the ratio $\frac{\hat{\alpha}}{(fA)}$. We note that even though we have used a second-order Legendre polynomial about the Cartesian $z$ axis, the dependence on $\beta_2$, the coefficient of the quadratic term, has completely disappeared. We obtain a closed-form solution
\begin{equation}
	\theta_c = \frac{1}{2}\arccos\left(\frac{-\hat{\alpha}}{6fA}\right),
\end{equation}
and, again, the only factor seemingly capable of determining $\theta_c$ is the ratio of mean strengths of the bulk stress-modifying mechanisms.

\begin{figure}[h!]
	\centering
	\includegraphics[totalheight=8cm]{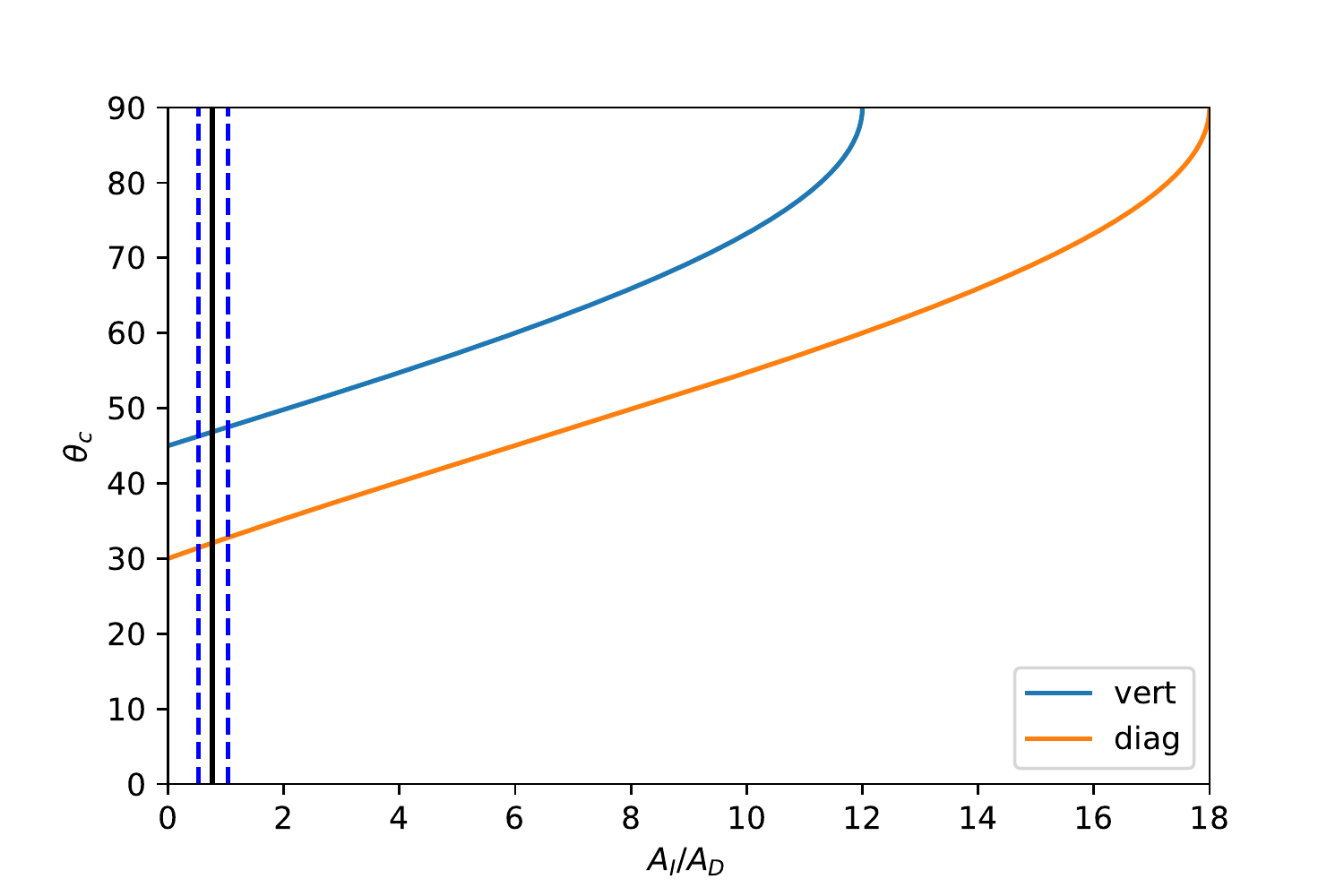}
	\caption{Critical angle selection as a function of the ratio of IIS strength to APF strength and the description of the interfacial geometry. \textbf{Black line:} mean estimate of the ratio. \textbf{Blue lines:} uncertainty in the fitting due to multiple experimental measurements as tracked by the \textit{uncertainties} Python package.}
	\label{famousplot}
\end{figure}

\paragraph{Qualitative results: interface relation, isotropy ratio, and $\theta_c$-selection.} 
Remarkably, in the case that the interface-shifts and the direction along which the stress is generated coincide perfectly, we find that depth-dependence completely factors out, and the only factor that can increase $\theta_c$ for a given interface relation is the ratio between $\hat{\alpha}$ and $fA$ (Figure \ref{famousplot}). However, the choice of interface relation is, itself, a major contributor to $\theta_c$ selection. This is unexpected and alarming, as all existing analyses of hydrodynamic stability for ion-irradiated thin films have considered only one interface relation each, often selecting the treatment of the interfaces merely on a basis of convenience. We have now shown that the relationship between the free and amorphous-crystalline interfaces may be the single largest missing component in an adequate description of irradiated thin films, especially as pertains to $\theta_c$ selection. 

\paragraph{Comparison of theory and experiment.}
We have obtained estimates of $fA$ and $\hat{\alpha}$ by fitting our theoretical steady-state stress to a small experimental data set. In combination with our analytical solutions for $\theta_c$, which depend only on the ratio $\frac{\hat{\alpha}}{fA}$ and the interface relation, we may compare theory and experiment. Figure \ref{famousplot} shows predictions for $\theta_c$ against these two quantities. It is immediately apparent that the assumption of vertically-translated interfaces leads to higher $\theta_c$ for all values of $\frac{\hat{\alpha}}{fA}$ when compared with the assumption of diagonally-translated interfaces that follow the ion track. Strikingly, the variation in our theoretical predictions is accounted for mostly by uncertainty about the correct shape of the interfaces rather than uncertainty about the ratio of the magnitudes of the bulk mechanisms. As we expect $\theta_c \approx 45^{\circ}$ for irradiation of Si by Ar$^+$ below about 1keV, it is clear that the experimentally-observed value of $\theta_c$ is captured within our range of theoretical predictions. Indeed, with the diagonal-translation case, we have low-end and high-end estimates of $\theta_c$ as $\sim 31^{\circ}$ and $\sim 33^{\circ}$ respectively. For the vertical-translation case, we have low-end and high-end estimates of $\theta_c$ as $\sim 46^{\circ}$ and $\sim 47.5^{\circ}$ respectively.

We expect that the true physical description of the interface relation is somewhere in between the idealized diagonal and vertical cases, as suggested by Figure 2 of \cite{norris-etal-SREP-2017}. This immediately prompts us to consider a more precise treatment of the interface relation. It is also worth noticing that although we have considered two idealizations of ion-induced stress modification in the vein of \cite{moreno-barrado-etal-PRB-2015}, these idealizations are strongly expected to be experimentally invalid. We have supposed that the intensity of each bulk mechanism is determined entirely by one ion track; this is in contrast to spatially-resolved models of power deposition in the style of \cite{bradley-PRB-2011b} which compute power deposition at a point as the integral of all incoming ions across the surface, which power deposition per ion following a Gaussian ellipse. Such an approach was necessary for the development of the earliest physically-descriptive theory of erosion and may also be necessary in the present hydrodynamic, stress-based context. With a more realistic treatment of the bulk mechanisms informed by ion-induced power deposition, it is possible that inhomogeneous stress modification will, indeed, influence $\theta_c$. This is certainly true if the two bulk mechanisms are not imagined to be identically distributed throughout the amorphous layer, as can already be seen from the calculations of the present work: the factoring-out of the depth-dependence term is entirely due to the identical distribution supposed for each mechanism, which is already a significant idealization.

\section{Discussion}
\subsection{Contributions and implications}
\paragraph{A high-generality two-factor model.} The present work provides the most general framework for hydrodynamic treatments of the linear stability of ion-irradiated surfaces to date, with arbitrary, substitution-ready quantities to describe the interface relation and depth-dependence profiles. The components of the stress tensor itself have also been left arbitrary throughout most of the calculations, permitting model refinements in the future.

\paragraph{On the importance of the interface relation.} 
Most analyses prior to \cite{Swenson_2018,evans-norris-JPCM-2022} have simply assumed an interface shape amenable to calculation, typically flat \cite{castro-cuerno-ASS-2012}, vertically-translated interfaces \cite{norris-PRB-2012-linear-viscous,norris-PRB-2012-viscoelastic-normal}, or diagonal \cite{moreno-barrado-etal-PRB-2015}, without considering the interface relation arbitrarily and carrying it throughout the entire calculation in unevaluated form. In doing so, and in considering two idealizations of this quantity in our analysis, we find that the interface relation can strongly influence stress modification and, therefore, critical angle selection. This immediately motivates further studies to refine it beyond flat, vertical, or diagonal, while the latter is admittedly a substantial improvement over the previous two, even if only because it permits physically-reasonable treatment of flux dilution and down-beam stress deposition.

\paragraph{On the role of inhomogeneous stress.} We have shown that when the axis along which stress is inserted agrees \textit{exactly} with the axis along which the interfaces translate, there is a cancellation effect and no depth-dependence along the downbeam coordinate are relevant (up to the third term of a Legendre polynomial in downbeam coordinates). This occurred in two such cases that we considered. However, we note that the present treatment, that of down-beam stress deposition along a vanishingly-thin ion-track is a significant idealization, as are the diagonally-translated interfaces that we have considered. In a more realistic treatment, we would not expect that all of the stress modification would act directly along a single ion-track, particularly in the nuclear stopping regime. Indeed, this was precisely the motivation for the Sigmund model of nuclear power deposition \cite{sigmund-PR-1969} and its subsequent, well-known analyses \cite{bradley-harper-JVST-1988,bradley-PRB-2011b}. We consider that in general, we should expect some stress-modification to occur ``off-axis", hence the inhomogeneity of depth-dependence may turn out to influence critical angle in all but certain idealized cases as those we have considered here. The unexpected, extreme importance of the interfacial geometry is cautionary: unjustified idealizations are hazardous.

\paragraph{Isotropic swelling and dwell-time considerations.}
It has been argued within our group's own previous work \cite{Swenson_2018,evans-norris-JPCM-2022} that one explanation for increased critical angles in some experimental systems may be due to enhanced dwell-times, which would grant the isotropic swelling mechanism increased strength relative to the anisotropic plastic flow mechanism, resulting in an enlarged $\frac{A_I}{A_D}$ ratio. As seen in (figure ref), such an enlarged ratio would, indeed, tend to increase the value of $\theta_c$. However, we must consider the form of the long-wave dispersion relation computed in the present work, a result from a composite model encompassing both bulk mechanisms. If the dispersion relation $\sigma$ are nondimensionalized by the scaling $\frac{h_0}{V}$ as was done in \cite{Swenson_2018,evans-norris-JPCM-2022}, which re-expresses the dimensionless growth-rates in terms of the mechanism strengths weighted by dwell time, both mechanisms receive the same scaling, and in the nondimensionalized ratio $\frac{\tilde{A}_I}{\tilde{A}_D}$, the dwell-time will cancel out. In the case of the arbitrary-swelling long-wave dispersion relation computed in \cite{evans-norris-JPCM-2022}, some dependence on dwell-time is retained but we expect that it will be too small to be relevant. Hence the longevity of a parcel of matter within the bulk is here revealed to be less significant than thought, at least under the assumption that APF and IIS are generated at the same locations in the film. We may therefore retain either the notion of dwell-time as relevant to critical angle selection, or we may retain the production of APF and IIS in the same regions of the film, \textit{but not both.} As an alternative, the present work appears to indicate that a strong contributing factor to the observed nonuniversality of critical angle $\theta_c$ may be found in refining the \textit{extremely important} amorphous-crystalline interface shape and, possibly, in a spatially-resolved treatment of the influence of power deposition on mechanism strength.

\paragraph{Interpretation of low-end critical angle estimate.}
Strikingly, the assumption of diagonally-translated interfaces, made in the work of \cite{moreno-barrado-etal-PRB-2015,Swenson_2018,evans-norris-JPCM-2022}, produces $\theta_c \approx 30^{\circ}$ in the absence of any swelling, a thoroughly wrong theoretical prediction, and requires about \textit{three times} as much swelling as the vertically-translated interfaces in order to achieve $\theta_c = 90^{\circ}$.

The finding that the diagonally-translated interfaces can produce $\theta_c \approx 30^{\circ}$ in the absence of swelling may have an interesting interpretation. It has been noted elsewhere \cite{hofsass-bobes-zhang-JAP-2016} that for very high-energy noble gas bombardment of monatomic substrates, $\theta_c$ may, indeed, reach as low as around $30^{\circ}$. Considering that the phenomenological form of the APF stress tensor has been borrowed from the electronic stopping regime, and insofar as it represents the ``pancake strain" resulting from intra-film melt cycles \cite{klaumunzer-etal-REDS-1989,trinkaus-ryazanov-PRL-1995-viscoelastic,trinkaus-NIMB-1998-viscoelastic,van-dillen-etal-PRB-2005-viscoelastic-model}, which we might anticipate would hinder the defect dynamics believed to be responsible for swelling \cite{steinbach-etal-PRB-2011,Swenson_2018}, a natural interpretation of this result might be that \textit{formation of the amorphous-crystalline interface attributable entirely to melts along the ion track, rather than the collision cascade, will naturally lead to the diagonal-translation case.} Pure APF and the diagonally-translated interfaces associated with high-energy irradiation, then, lead to $\theta_c = 30^{\circ}$ as a special limit. This creates the possibility of a natural bridge between the electronic stopping regime, characterized by pure diagonal translation and energy deposition primarily along ion tracks with swelling due to defect clustering largely suppressed by the melt cycle, and the nuclear stopping regime, characterized by interface translation perhaps closer to vertical and the possibility of well-developed defect dynamics.

%Note also that for 250eV Ar+ on Si, $\frac{h_0}{V}$ (dwell-time) is relatively large compared to other energies (see Figure 2 in Swenson-Norris 2018). We therefore consider that the ``small cross-terms" limit should be of interest in general systems. 

\paragraph{Relationship with existing depth-dependence model.} 
%\begin{enumerate}
%	\item Discussion of Moreno-Barrado model
%	\item Discussion of Hofs\"{a}ss and modified CFF
%	\item Discussion of Norris PRB 2012b vs EBF (Spaniard) models, re: the resulting velocity profile and comparison with gravity-driven flow/effective body forces. Stress-driven flow versus body-force driven flow; \textit{quadratic versus linear velocity profile in the bulk.}
%\end{enumerate}

The present work naturally evokes comparison to a previous discussion of depth-dependent stress due to \cite{moreno-barrado-etal-PRB-2015}. However, several substantial differences exist. First, the fundamental form of the stress tensor, even before depth-dependence considerations, is different, and differently motivated. The present work bases its analysis on the phenomenological use of a stress tensor found in the nuclear stopping regime work of \cite{van-dillen-etal-APL-2001-colloidal-ellipsoids,van-dillen-etal-APL-2003-colloidal-ellipsoids,van-dillen-etal-PRB-2005-viscoelastic-model,otani-etal-JAP-2006,george-etal-JAP-2010,norris-PRB-2012-linear-viscous}, which draws analogy with the ``pancake strain" model of anisotropic stress in irradiated films at higher, electronic stopping regime energies. On the other hand, that of \cite{moreno-barrado-etal-PRB-2015} could be considered as an alternative phenomenological model seeking analogy with classical hydrodynamic instabilities \cite{chandrasekhar-book-2013}, such as B\'{e}nard-Marangoni convection. (If anything, the existence of two such mathematically valid phenomenological models giving different predictions for the same physical system is an argument in favor of leveraging phenomenological models only to the extent necessary to develop true physically-grounded models---and then abandoning the phenomenological models.) It is therefore immediately unsurprising that we have obtained different results. Additionally, the present model allows an additional stabilizing mechanism via isotropic swelling, whereas that of most other hydrodynamic analyses of irradiated thin films consider only anisotropic stress (although not necessarily APF) due to the ion beam. For a more detailed discussion on the differences between our model and that of \cite{moreno-barrado-etal-PRB-2015}, we refer the reader to the Appendix.

\subsection{Open questions and future work}
\paragraph{Refinement of interface modeling.}
Throughout the literature, three main assumptions have been made about the relationship between the upper and lower interfaces: a flat lower interface as well as vertical and diagonal displacements of the lower from the upper have been considered and incorporated into existing analyses. As discussed elsewhere \cite{NorrisAziz_predictivemodel}, a major outstanding problem in the field is not merely achieving good agreement between theory and experiment (where fitting data to, e.g., anisotropic Kuramoto-Sivashinsky can lead to impressive results even in the nonlinear regime without suggesting underlying mechanisms), but \textit{identifying the underlying mechanisms leading to good agreement}: that is, a ``bottom-up" approach to modeling the appropriate physics, and developing models which are precisely as simple or as complicated as needed (but no simpler nor more complicated). The present results suggest that a significant and hitherto-unappreciated component of any analysis is the correct analytical treatment of the interface relation, without which any apparent agreement between theory and experiment must be regarded as possibly \textit{coincidental}.

\paragraph{Underlying physics in regime of dominant nuclear stopping.}
In the present work, we have found that variations up to quadratic order in the strength of the two theorized bulk mechanisms lead to no changes in $\theta_c$. We note that such variation may still be important for wavelength selection, which is best studied outside of the context of a long-wave analysis as we have conducted here. This finding is entirely contingent upon having supposed that anisotropic plastic flow and isotropic swelling occur simultaneously and vary identically throughout the film depth. This is a fundamental assumption about the physical origins of each mechanism--- and a na\"{i}ve one. The physical origin of anisotropic plastic flow, and its apparent good agreement as a theoretical model with experimental results, is a curiosity currently lacking a comprehensive explanation. It is likely that isotropic swelling is driven by either in-film defect dynamics or ion-implantation disrupting the energetically-preferred lattice in Si. Such explanations have led to good agreement between theory and experiment elsewhere \cite{ishii-etal-JMR-2014},\cite{chan-chason-JVSTA-2008}, while fewer good explanations exist for anisotropic plastic flow in low-energy irradiation. If it is shown that the anisotropic and isotropic components of ion-induced stress do not, in fact, vary together, our finding that depth-dependence cannot affect $\theta_c$ is immediately reversed, although it is possible that the dependence may still be \textit{weak}. Nonetheless, because we have derived our results for fully arbitrary depth-dependence profiles, the present work will be easily adapted to any future findings on this matter.

\paragraph{Fully resolved spatial variation.} In the present work, we have studied idealized cases where all spatial variation occurs along a downbeam path of vanishing crossbeam thickness. We have found that this idealization leads to no influence of spatial variation on $\theta_c$ selection, which is at first surprising. However, this result may have a subtle interpretation. We note that in the Bradley-Harper model of erosion \cite{bradley-harper-JVST-1988,bradley-PRB-2011b}, the well-known Bradley-Harper instability only becomes apparent when considering fully spatially-resolved collision cascades: an attempt at an erosive model similar to Bradley-Harper but considering deposition only along the downbeam axis \textit{would naturally obscure this classical instability} which originates from differential power deposition at ``valleys" and ``hilltops". Hence one interpretation of the present results is that a spatially-resolved model, analogous to Bradley-Harper but occurring in the amorphous bulk, will be required to fully capture the influence of spatially-varying bulk mechanisms on $\theta_c$ selection. 

\paragraph{Experimental parameter estimation.}
As a basis for our comparison between theory and experiment, we have attempted to estimate the values of two parameters from a small wafer-curvature data set from experiments involving 250eV irradiation of Si by Ar$^+$ by fitting the angle-dependent data to the component of our steady-state stress tensor associated with in-plane stress. Reasonable agreement was found between the data and the theoretical in-plane stress model, and parameter estimates appear to be in line with those from elsewhere in the literature. Nonetheless, approximation by the steady-state stress is imperfect, and the data set was small. Further angle-dependent wafer curvature measurements leading to separation of isotropic and anisotropic stress components for various materials would be of great benefit in connecting theoretical and experimental results, especially across a variety of energies, projectiles and targets.

\section*{Acknowledgments}
We gratefully acknowledge support from the National Science Foundation through DMS-1840260.

\appendix
\section{Details of linear stability analysis}
Here, we provide a complete solution under a relatively simple, but general, case that
\begin{equation}
	\tau = \tau(z;g,h), \alpha = \alpha(z;g,h);
\end{equation}
it will be seen that the generalized case in the main text is easily obtained from this.
\subsection{General equations}
Before beginning the analysis, it is convenient to compute in advance the following, as described in the main text:
\begin{equation}
	\alpha(z;0 + \epsilon g_{1},h_0 + \epsilon h_{1}) \approx \alpha_0(z;0,h_0) + \epsilon[\alpha_{g}g_{1} + \alpha_{h}h_{1}]|_{(z;0,h_0)},
\end{equation}
and
\begin{equation}
	\tau(z; 0+\epsilon g_{1}, h_0 + \epsilon h_{1}) \approx \tau(z;0,h_0) + \epsilon[\tau_h|_{(z;0,h_0)}h_1 + \tau_g|_{(z;0,h_0)}g_1].
\end{equation}
Here, we also note the general momentum balance equations in the bulk for the reader's reference:
\begin{equation}
	\begin{gathered}
		-p_x + \eta\{2u_{xx} + u_{yy} + v_{xy} + u_{zz} + w_{xz}\} = 2fA\eta\{D_{11}\frac{d \tau}{d x} + D_{13}\frac{d \tau}{d z} \} \\ 
		-p_y + \eta\{u_{xy} + v_{xx} + 2v_{yy} + v_{zz} + w_{yz}\} = 2fA\eta\frac{d \tau}{d y} \\
		-p_z + \eta\{u_{xz} + v_{yz} + w_{xx} + w_{yy} + 2w_{zz}\} = 2fA\eta\{D_{31}\frac{d \tau}{d x} + D_{33}\frac{d \tau}{d z} \}
	\end{gathered}
\end{equation}
We then naturally have
\begin{equation}
	\begin{gathered}
		-p_x + \eta\{2u_{xx} + u_{yy} + v_{xy} + u_{zz} + w_{xz}\} = 2fA\eta\{D_{11}(\tau_gg_x + \tau_hh_x ) + D_{13}\tau_z \} \\ 
		-p_y + \eta\{u_{xy} + v_{xx} + 2v_{yy} + v_{zz} + w_{yz}\} = 2fA\eta(\tau_gg_y + \tau_hh_y) \\
		-p_z + \eta\{u_{xz} + v_{yz} + w_{xx} + w_{yy} + 2w_{zz}\} = 2fA\eta\{D_{31}(\tau_gg_x + \tau_hh_x ) + D_{33}\tau_z \},
	\end{gathered}
\end{equation}
from which we will proceed. We note that we have not assumed standard incompressibility in order to simplify the above momentum balance equations (i.e., allowing cancellation of several mixed partials). We will consider only the projected down-beam direction, as this is sufficient for a study of critical angle selection; hence we strip $y$-dependence from the following calculations.

\subsubsection{Steady state and first expansion: $\epsilon$-small perturbations} Following the application of the steady-state and translation-invariance assumptions ($\frac{\partial}{\partial x} \to 0, \frac{\partial}{\partial t} \to 0$), linearization about small perturbations produces the steady-state equations
\begin{equation}
	\begin{gathered}
		\frac{\partial}{\partial z}(\rho_{0}w_0) = 0 \\
		\eta u_{0zz} = 2\eta fAD_{13}\tau_z\\
		-p_{0z} + 2\eta w_{0zz} = 2\eta fAD_{33}\frac{\partial \tau}{\partial z} \\
		\rho_0 = \frac{\rho^*}{1+\Delta_0} \\
		w_0 \Delta_{0z} = \alpha(z;0,h_0).
	\end{gathered}
\end{equation}
At z=0, we have
\begin{equation}
	\begin{gathered}
		\Delta_0 = 0 \\
		u_0 = 0 \\
		w_0 = V,
	\end{gathered}
\end{equation}
representing the steady-state equations of the no-slip and no-penetration conditions respectively in the downward-translating frame. At z=$h_0$, we have
\begin{equation}
	\begin{gathered}
		0 = w_0 - V\frac{\rho^*}{\rho_0} \\
		u_{0,z} = 2fA\tau(z;0,h_0)D_{13} \\
		p_0 = 2\eta\{w_{0z} - fA\tau(z;0,h_0)D_{33}\},
	\end{gathered}
\end{equation}
where the first equation is the steady-state equation of the modified kinematic condition, and the second two equations are due to the steady-state stress balance $\textbf{T}_{0}\cdot \hat{n}_{0} = 0$ at the upper interface. We also find the linearized equations at $O(\epsilon)$ as
\begin{equation}
	\begin{gathered}
		\rho_{1t} + \rho_{0}u_{1x} + \rho_{0z}w_1 + \rho_{0}w_{1z} + \rho_{1z}w_0 + \rho_{1}w_{0z} = 0 \\
		-p_{1x} + \eta\{2u_{1xx} + u_{1zz} + w_{1xz}\} =  2fA\eta\{D_{11}[\tau_{h}h_{1x} + \tau_{g}g_{1x} ] + D_{13}[\tau_{zh}h_{1} + \tau_{zg}g_{1}]\} \\
		-p_{1z} + \eta\{w_{1xx} + 2w_{1zz} + u_{1xz} \} = 2fA\eta\{D_{31}[\tau_{h}h_{1x} + \tau_{g}g_{1x}] + D_{33}[\tau_{zh}h_{1} + \tau_{zg}g_{1}]\} \\
		\rho_{1} = \frac{-\rho^*\Delta_{1}}{(1+\Delta_0)^2} \\
		\Delta_{1t} + u_0\Delta_{1x} +  w_{0}\Delta_{1z} + w_{1}\Delta_{0z} = [\alpha_{g}g_1 + \alpha_h h_{1}]|_{(z;0,h_0)}
	\end{gathered}
\end{equation}
in the bulk, representing conservation of mass and momentum; the equation of state; and the advection of the volume-change respectively. Now at $z=0$,
\begin{equation}
	\begin{gathered}
		\Delta_{0,z}(z;0,h_0)g_{1} + \Delta_{1}(z;0,h_0) = 0 \\
		u_{0z}g_{1} + u_{1} = 0 \\
		w_{0z}g_{1} + w_{1} = 0.
	\end{gathered} 
\end{equation}
At $z=h_0$,
\begin{equation}
	\begin{gathered}
		h_{1t} = w_1 -u_0h_{1x} + h_1w_{0z} + \frac{V\rho^*}{\rho_0^2}\{\rho_{0z}h_1 + \rho_1\} \\
		2\eta\{\frac{1}{2}(u_{1z} + w_{1x}) - fA[\tau_{h}h_1 + \tau_{g}g_1]D_{13}\} - h_{1x}T_0^{11} = 0 \\
		-p_1 + 2\eta\{w_{1z} - fA[\tau_{h}h_1 + \tau_{g}g_1]D_{33}\} + T_{0z}^{33} = 0,
	\end{gathered}
\end{equation}
where the first equation is due to the linearization of the kinematic condition and the second two are due to the linearization of the stress balance,
\begin{equation}
	\textbf{T}_0 \cdot \hat{n}_1 + \left[\frac{\partial \textbf{T}_0}{\partial z}\cdot h_1 + \textbf{T}_1 \right]\cdot \hat{n}_0 = \vec{0}.
\end{equation}
Here, we have $\hat{n}_0 = <0, 0, 1>$ and $\hat{n}_1 = <-h_{1x}, -h_{1y}, 0>$. $T_0^{11}$ denotes the upper-left component of steady-state stress tensor $\textbf{T}_0$, and $T_0^{33}$ denotes the bottom-right component. The component indices are denoted as superscripts to distinguish them from the subscripts which elsewhere denote terms in the expansion.

\subsubsection{Second expansion: small perturbative wavenumber $k$}
\noindent Expanding the scalar fields above in normal modes of the form
\begin{equation}
	F(x,z,t) = \tilde{F}(z)\exp(\sigma t + ikx)
\end{equation}
and the interfaces as, say,
\begin{equation}
	\begin{gathered}
		h_1(x,t) = \tilde{h}_1\exp(\sigma t + ikx) \\
		g_1(x,t) = \tilde{g}_1\exp(\sigma t + ikx).
	\end{gathered}
\end{equation}
We now seek the long-wave linear dispersion relation, since we anticipate that pattern formation in this unary material material should be governed by a so-called Type II bifurcation (cite cite). This prompts the expansion of the dispersion relation $\sigma \approx 0$ in small wavenumber k as
\begin{equation}
	\sigma = 0 + k \sigma_1 + k^2 \sigma_2 + O(k^3),
\end{equation}
and we obtain the following systems at each order in $k$.
\paragraph{At O(1):}
\begin{equation}
	\begin{gathered}
		\rho_{0z}\tilde{w}_{10} + \rho_0\tilde{w}_{10}' + \tilde{\rho}_{10}'w_{0} + \tilde{\rho}_{10}w_{0z} = 0 \\
		\tilde{u}_{10}'' = 2fAD_{13}[\tau_{zh}\tilde{h}_{1} + \tau_{zg}\tilde{g}_{1}] \\
		-\tilde{p}_{
			10}' + 2\eta w_{10}'' = 2fA\eta\{D_{33}[\tau_{zh}\tilde{h}_{1} + \tau_{zg}\tilde{g}_{1}]\} \\
		\tilde{\rho}_{10} =  \frac{-\rho^* \tilde{\Delta}_{10}}{(1+\Delta_0)^2} \\
		w_0\tilde{\Delta}_{10}' + \tilde{w}_{10}\Delta_{0z} = [\frac{\partial \alpha}{\partial g}\tilde{g}_{1} + \frac{\partial \alpha}{\partial h}\tilde{h}_{1}].
	\end{gathered}
\end{equation}
At $z=0$,
\begin{equation}
	\begin{gathered}
		\Delta_{0z}g_1 + \tilde{\Delta}_{10} = 0 \\
		\tilde{u}_{10} + u_{0z}\tilde{g}_1 = 0 \\
		\tilde{w}_{10} + w_{0z}\tilde{g}_1 = 0
	\end{gathered}
\end{equation}
At $z=h_0$,
\begin{equation}
	\begin{gathered}
		\tilde{w}_{10} + \tilde{h}_1 w_{0z} + \frac{V\rho^*}{\rho_0^2}\{\rho_{0z}\tilde{h}_{1} + \tilde{\rho}_{10}\} = 0 \\
		2\eta\{\frac{\tilde{u}_{10}'}{2} - fA[\tau_h\tilde{h}_{1} + \tau_g\tilde{g}_{1}]D_{13}\} = 0 \\
		-\tilde{p}_{10} + 2\eta\{\tilde{w}_{10}' - fA[\tau_h\tilde{h}_{1} + \tau_g\tilde{g}_{1}]D_{33}\} + T_{0z}^{33} = 0
	\end{gathered}
\end{equation}

\paragraph{At O(k):}
\begin{equation}
	\begin{gathered}
		\sigma_1 \tilde{\rho}_{10} + i\rho_{0}\tilde{u}_{10} + \rho_{0z}\tilde{w}_{11} + \rho_0\tilde{w}_{11}' + \tilde{\rho}_{11}'w_0 + \tilde{\rho}_{11}w_{0z} = 0 \\
		-i\tilde{p}_{10} + \eta \{ \tilde{u}_{11}'' + i\tilde{w}_{10}' \}= 2fA\eta i\{D_{11}[\tau_h \tilde{h}_{1} + \tau_g \tilde{g}_{1} ] \} \\
		-\tilde{p}_{11}' + \eta \{ 2\tilde{w}_{11}'' + i\tilde{u}_{10}'\} = 2fA\eta i\{D_{31}[\tau_h \tilde{h}_{1} + \tau_g \tilde{g}_{1} ] \} \\
		\tilde{\rho}_{11} = \frac{-\rho^*\tilde{\Delta}_{11}}{[1+\Delta_0]^2} \\
		\sigma_1 \tilde{\Delta}_{10} + iu_0\tilde{\Delta}_{10} + w_0\tilde{\Delta}_{11}' + \tilde{w}_{11}\Delta_{0z} = 0
	\end{gathered}
\end{equation}
At $z=0$,
\begin{equation}
	\begin{gathered}
		\tilde{\Delta}_{11} = 0 \\
		\tilde{u}_{11} = 0 \\
		\tilde{w}_{11} = 0
	\end{gathered}
\end{equation}
At $z=h_0$,
\begin{equation}
	\begin{gathered}
		\sigma_{1}\tilde{h}_{1} = \tilde{w}_{11} -u_0i\tilde{h}_1 + \frac{V \rho^*}{\rho_0^2}\tilde{\rho}_{11} \\
		2\eta\{\frac{1}{2}[\tilde{u}_{11}' + i\tilde{w}_{10}]\} - i \tilde{h}_{1}T_0^{11} = 0 \\
		-\tilde{p}_{11} + 2\eta \tilde{w}_{11}' = 0
	\end{gathered}
\end{equation}

\paragraph{At O($k^2$):}
\begin{equation}
	\begin{gathered}
		\sigma_{1}\tilde{\rho}_{11} + \sigma_{2} \tilde{\rho}_{10} + i\rho_{0}\tilde{u}_{11} + \rho_{0z} \tilde{w}_{12} + \rho_0 \tilde{w}_{12}' + \tilde{\rho}_{12}' w_0 + \tilde{\rho}_{12} w_{0z} = 0 \\
		-i \tilde{p}_{11} + \eta\{-2\tilde{u}_{10} + \tilde{u}_{12}'' + i\tilde{w}_{11}'\} = 0 \\
		-\tilde{p}_{12}' + \eta\{-\tilde{w}_{10} + 2\tilde{w}_{12}'' + i\tilde{u}_{11}' \} = 0 \\
		\tilde{\rho}_{12} = \frac{-\rho^*\tilde{\Delta}_{12}}{[1+\Delta_0]^2} \\
		\sigma_{1}\tilde{\Delta}_{11} + \sigma_{2}\tilde{\Delta}_{10} + iu_0\tilde{\Delta}_{11} + w_{0}\tilde{\Delta}_{12}' + \tilde{w}_{12}\Delta_{0z} = 0
	\end{gathered}
\end{equation}
at z=0,
\begin{equation}
	\begin{gathered}
		\tilde{\Delta}_{12} = 0 \\
		\tilde{u}_{12} = 0 \\
		\tilde{w}_{12} = 0,
	\end{gathered}
\end{equation}
and at z=$h_0$,
\begin{equation}
	\begin{gathered}
		\sigma_{2} \tilde{h}_{1} = \tilde{w}_{12} + \frac{V \rho^*}{\rho_0^2} \tilde{\rho}_{12} \\ 
		\eta\{\tilde{u}_{12}' + i\tilde{w}_{11}\} = 0 \\
		-\tilde{p}_{12} + 2\eta\tilde{w}_{12}' = 0.
	\end{gathered}
\end{equation}

\subsubsection{Third expansion: small swelling rate $\hat{\alpha}$}
It quickly becomes apparent that a fully-general solution of the above equations, resulting in the closed-form linear dispersion relation, is not mathematically tractable. In particular, by combining mass conservation, age-advection and the equation of state, we may obtain the ordinary differential equation,
\begin{equation}
	(1+\Delta_0)\Delta_{0z} = \frac{\rho^*}{a_1}\alpha(z;0,h_0).
\end{equation}
Integration in $z$ leads immediately to
\begin{equation}
	\Delta_0^2 + \Delta_0 - \frac{\rho^*}{a_1}\left( \int_0^z\alpha(\tilde{z};0,h_0)d\tilde{z} + a_2\right) = 0,
\end{equation}
and we obtain
\begin{equation}
	\Delta_0 = \frac{-1 + \sqrt{1 + \frac{4}{V}\int_{0}^z\alpha(\tilde{z};0,h_0)d\tilde{z}}}{2}
\end{equation}
when boundary conditions are considered. A solution may then be determined for $w_0$ in terms of $\Delta_0$. However, these solutions, as they appear in the $O(\epsilon)$ equations, result in a system of ordinary differential equations with complicated, nonlinear coefficients, even for $\alpha$ linear in $z$.

Bearing in mind that we expect the swelling rate in the traveling frame to be fairly small (that is, there isn't much swelling prior to a parcel of matter being sputtered away), and observing from previous results (cite Evans-Norris) that the effect of even large swelling rates is highly self-similar at all wave numbers, and uniformly stabilizing for long waves, we consider a second expansion in a small mean swelling rate, $\hat{\alpha}$. Hence we take
\begin{equation}
	\begin{gathered}
		\alpha = \alpha_{0} + \hat{\alpha}\alpha_{1} + ... \\
		\rho_{0} = \rho_{00} + \hat{\alpha}\rho_{01} + ...\\
		... \\
		\sigma_{1} = \sigma_{10} + \hat{\alpha}\sigma_{11} + ... \\
		\tilde{\rho}_{10} = \tilde{\rho}_{100} + \hat{\alpha}\tilde{\rho}_{111} + ... \\
		...
	\end{gathered}
\end{equation}
Now, because the equations for the leading order terms are obvious from the above (simply by appending a ``0" to the subscript of each term), we shall only write out explicitly the equations expanded in $\hat{\alpha}$. Because the stress tensor associated with anisotropic plastic flow is assumed, here, to be independent of swelling rate, there is no need for us to expand $\tau \to \tau_{0} + \hat{\alpha}\tau_{1}$. In a more careful exploration of the interplay between volumization and stress-free strain through local changes in the density, such an expansion would be required.

\paragraph{Steady state at O($\hat{\alpha}$):}
\begin{equation}
	\begin{gathered}
		\frac{\partial}{\partial z}(\rho_{00}w_{01} + \rho_{01}w_{00}) = 0 \\
		\eta u_{01zz} = 0 \\
		-p_{01z} + 2\eta w_{01zz} = 0 \\
		\rho_{01} = \frac{-\rho^* \Delta_{01}}{(1+\Delta_{00})^2} \\
		w_{00}\Delta_{01z} + w_{01}\Delta_{00z} = \alpha_{1}(z;0,h_0).
	\end{gathered}
\end{equation}
At z=0:
\begin{equation}
	\begin{gathered}
		\Delta_{01} = 0 \\
		u_{01} = 0 \\
		w_{01} = 0.
	\end{gathered}
\end{equation}
At z=$h_0$,
\begin{equation}
	\begin{gathered}
		w_{01} + V\rho^*\frac{\rho_{01}}{\rho_{00}^2} = 0 \\
		u_{01z} = 0 \\
		-p_{01} + 2\eta w_{01z} = 0.
	\end{gathered}
\end{equation}

\paragraph{At O($\hat{\alpha}$):}
\begin{equation}
	\begin{gathered}
		\rho_{00z}\tilde{w}_{101} + \rho_{01z}\tilde{w}_{100} + \rho_{00}\tilde{w}_{101z} + \rho_{01}\tilde{w}_{100z} + \tilde{\rho}_{100z}w_{01} + \tilde{\rho}_{101z}w_{00} + \tilde{\rho}_{100}w_{01z} + \tilde{\rho}_{101}w_{00z} = 0 \\
		\tilde{u}_{101zz} = 0 \\
		-\tilde{p}_{101z} + 2\eta \tilde{w}_{101zz} = 0 \\
		w_{00}\tilde{\Delta}_{101z} + w_{01}\tilde{\Delta}_{100z} + \tilde{w}_{100}\Delta_{01z} + \tilde{w}_{101}\Delta_{00z} = [\frac{\partial \alpha_1}{\partial g}\tilde{g}_{1} + \frac{\partial \alpha_1}{\partial h}\tilde{h}_{1}] \\
		\tilde{\rho}_{101} = \frac{-\rho^*\{\Delta_{00}\tilde{\Delta}_{101} + \tilde{\Delta}_{101} - 2\Delta_{01}\tilde{\Delta}_{100}  \}}{(1+\Delta_{00})^3} 
	\end{gathered}
\end{equation}
At z=0:
\begin{equation}
	\begin{gathered}
		\Delta_{01z}\tilde{g}_{1} + \tilde{\Delta}_{101} = 0 \\
		\tilde{u}_{101} + u_{01z}g_{1} = 0 \\
		\tilde{w}_{101} + w_{01z}g_{1} = 0.
	\end{gathered}
\end{equation}
At z=$h_{0}$:
\begin{equation}
	\begin{gathered}
		\tilde{w}_{101} + \tilde{h}_{1}w_{01z} + V\rho^*\frac{\{ \rho_{00}(\rho_{01z}h_{1} + \tilde{\rho}_{101}) -2\rho_{01}(\rho_{00z}h_{1} + \tilde{\rho}_{100})\}}{\rho_{00}^3} = 0\\
		\tilde{u}_{101}' = 0 \\
		-\tilde{p}_{101} + 2\eta \tilde{w}_{101}' + T_{01z}^{33}= 0.
	\end{gathered}
\end{equation}

\paragraph{At O($k\hat{\alpha}$):}
\begin{equation}
	\begin{gathered}
		\sigma_{10}\tilde{\rho}_{101} + \sigma_{11}\tilde{\rho}_{100} + i[\rho_{00}\tilde{u}_{101} + \rho_{01}\tilde{u}_{100}] + [\rho_{00z}\tilde{w}_{111} + \rho_{01z}\tilde{w}_{110}] \\ + [\rho_{00}\tilde{w}_{111}' + \rho_{01}\tilde{w}_{110}'] + [\tilde{\rho}_{110}'w_{01} + \tilde{\rho}_{111}'w_{00}] + [\tilde{\rho}_{110}w_{01z} + \tilde{\rho}_{111}\tilde{w}_{00z}] = 0 \\
		-i\tilde{p}_{101} + \eta\{\tilde{u}_{111}'' + i\tilde{w}_{101}'\} = 0 \\
		-\tilde{p}_{111}' + \eta\{2\tilde{w}_{111}'' + i\tilde{u}_{101}'\} = 0 \\
		\sigma_{10}\tilde{\Delta}_{101} + \sigma_{11}\tilde{\Delta}_{100} + i[u_{00}\tilde{\Delta}_{101} + u_{01}\tilde{\Delta}_{100}] +  w_{00}\tilde{\Delta}_{111}' + w_{01}\tilde{\Delta}_{110}' + \tilde{w}_{110}\Delta_{01z} + \tilde{w}_{111}\Delta_{00z} = 0 \\
		\tilde{\rho}_{111} = \frac{-\rho^*\{\Delta_{00}\tilde{\Delta}_{111} + \tilde{\Delta}_{111} - 2\Delta_{01}\tilde{\Delta}_{110}  \}}{(1+\Delta_{00})^3} 
	\end{gathered}
\end{equation}
At z=0:
\begin{equation}
	\begin{gathered}
		\tilde{\Delta}_{111} = 0 \\
		\tilde{u}_{111} = 0 \\
		\tilde{w}_{111} = 0.
	\end{gathered}
\end{equation}
At z=$h_{0}$:
\begin{equation}
	\begin{gathered}
		\sigma_{11}\tilde{h}_{1} = \tilde{w}_{111} -u_{01}i\tilde{h}_1+ \frac{V\rho^*[\rho_{00}\tilde{\rho}_{111} - 2\rho_{01}\tilde{\rho}_{110}]}{\rho_{00}^3} \\
		-i\tilde{h}_1T_{01}^{11} + \eta\{\tilde{u}_{111}' + i\tilde{w}_{101}\} = 0 \\
		-\tilde{p}_{111} + 2\eta \tilde{w}_{111}' = 0
	\end{gathered}
\end{equation}

\paragraph{At O($k^2\hat{\alpha}$):}
\begin{equation}
	\begin{gathered}
		\sigma_{10}\tilde{\rho}_{111} + \sigma_{11}\tilde{\rho}_{110} + [\sigma_{20}\tilde{\rho}_{101} + \sigma_{21}\tilde{\rho}_{100}] + i[\rho_{00}\tilde{u}_{111} + \rho_{01}\tilde{u}_{110}] \\ + [\rho_{00z}\tilde{w}_{121} + \rho_{01z}\tilde{w}_{120}] + [\rho_{00}\tilde{w}_{121}' + \rho_{01}\tilde{w}_{120}'] + [\tilde{\rho}_{120}'w_{01} + \tilde{\rho}_{121}'w_{00}] + [\tilde{\rho}_{120}w_{01z} + \tilde{\rho}_{121}w_{00z}] = 0 \\
		-i\tilde{p}_{111} + \eta\{-2\tilde{u}_{101} + \tilde{u}_{121}'' + i\tilde{w}_{111}'\} = 0 \\
		-\tilde{p}_{121}' + \eta\{-\tilde{w}_{101} + 2\tilde{w}_{121}'' + i\tilde{u}_{111}'\} = 0\\
		\sigma_{10}\tilde{\Delta}_{111} + \sigma_{11}\tilde{\Delta}_{110} + \sigma_{20}\tilde{\Delta}_{101} + \sigma_{21}\tilde{\Delta}_{100} + i[u_{00}\tilde{\Delta}_{111} + u_{01}\tilde{\Delta}_{110}] + w_{00}\tilde{\Delta}_{121}' + w_{01}\tilde{\Delta}_{120}' + \tilde{w}_{120}\Delta_{01z} + \tilde{w}_{121}\Delta_{00z} = 0 \\
		\tilde{\rho}_{121} = \frac{-\rho^*\{\Delta_{00}\tilde{\Delta}_{121} + \tilde{\Delta}_{121} - 2\Delta_{01}\tilde{\Delta}_{120}  \}}{(1+\Delta_{00})^3}
	\end{gathered}
\end{equation}
At z=0:
\begin{equation}
	\begin{gathered}
		\tilde{\Delta}_{121} = 0 \\
		\tilde{u}_{121} = 0 \\
		\tilde{w}_{121} = 0.
	\end{gathered}
\end{equation}
At z=$h_0$:
\begin{equation}
	\begin{gathered}
		\sigma_{21}\tilde{h}_1 = \tilde{w}_{121} + V\rho^*\frac{\rho_{00}\tilde{\rho}_{121} - 2\rho_{01}\tilde{\rho}_{120} }{\rho_{00}^3} \\
		\tilde{u}_{121}' + i\tilde{w}_{111} = 0 \\
		-\tilde{p}_{121} + 2\eta\tilde{w}_{121}' = 0 \\
	\end{gathered}
\end{equation}

\subsection{Solution}
Because of the lengthy calculations and the significance of the claims made in the present work, we include here the details of calculation leading to the theoretical steady-state stress and the dispersion relation.
\paragraph{Steady state solutions at O(1).} Observe that $w_{00}\Delta_{00z} = 0$ implies that either $w_{00} = 0$ or $\Delta_{00z} = 0$. If $w_{00} = 0$, we have no means of determining either $\Delta_{00}$ or $\rho_{00}$, but this solution appears physically unreasonable so we disregard it. On the other hand, $\Delta_{00z} = 0$ implies $\Delta_{00} = c_0$, and the first boundary condition at $z=0$ implies $c_0 =0$. Then $\rho_{00} = \rho^*$, leading to $w_{00z}$ = 0 in the steady-state mass conservation equation in the bulk. Then $w_{00}$ is a constant, and the boundary conditions at both interfaces imply $w_{00} = V$, the erosion rate. Steady-state momentum conservation in the bulk implies that
\begin{equation}
	\begin{gathered}
		u_{00z} = 2fAD_{13}\tau + c_1 \to u_{00} = 2fAD_{13}\int_{0}^{z}\tau d\tilde{z} + c_1z + c_2 \\
		-p_{00} = 2\eta fAD_{33}\tau + c_3.
	\end{gathered}
\end{equation}
Observe that the third boundary condition at the upper interface implies that $c_3 = 0$, and the second boundary conditions at both the upper and lower interfaces imply $c_1 = c_2 = 0$. Thus we have solution set
\begin{equation}
	\begin{gathered}
		\Delta_{00} = 0, \rho_{00} = \rho^*, w_{00} = V, \\
		u_{00} = 2fAD_{13}\int_{0}^z\tau d\tilde{z}, p_{00} = -2\eta fA\tau D_{33}.
	\end{gathered}
\end{equation}
\paragraph{Steady state solutions at O($\hat{\alpha}$).}
From the steady-state bulk equation for advection of volumization, including the knowledge that $\Delta_{00} = 0$ and $w_{00} = V$ from the previous calculations (regardless of the depth dependence profile, which has been left general), we have
\begin{equation}
	\Delta_{01z} = \frac{1}{V}\alpha_{1} \to \Delta_{01} = \frac{1}{V}\int_0^z \alpha_1 d\tilde{z} + c_4,
\end{equation}
and the first boundary condition at $z=0$ implies $c_4=0$. This leads immediately to the solution for $\rho_{01}$,
\begin{equation}
	\rho_{01} = \frac{-\rho^*}{V}\int_0^z \alpha_1 d\tilde{z}.
\end{equation}
Knowing $\rho_{01}$ allows us to determine $w_{01}$ from the steady state bulk mass conservation equation, and we obtain $\rho^* w_{01} + \rho_{01}V = c_5$, implying
\begin{equation}
	w_{01} = \int_0^z\alpha_1 d\tilde{z} + c_5.
\end{equation}
Consideration of the boundary conditions leads to $c_5 = 0$. Steady-state momentum conservation in the bulk leads to $-p_{01} + 2\eta w_{01z} = 0$, hence
\begin{equation}
	p_{01} = 2\eta \alpha_1.
\end{equation}
Finally, the other steady-state momentum conservation equation in the bulk leads to $u_{01}$ being at most linear in $z$, and the two homogeneous boundary conditions force $u_{01} = 0$. We therefore have the solution set
\begin{equation}
	\begin{gathered}
		\Delta_{01} = \frac{1}{V}\int_0^z \alpha_1 d\tilde{z}, \rho_{01} = \frac{-\rho^*}{V}\int_0^z \alpha_1 d\tilde{z}, \\ w_{01} = \int_0^z\alpha_1 d\tilde{z}, u_{01} = 0, p_{01} = 2\eta \alpha_1.
	\end{gathered}
\end{equation}

\paragraph{General strategy.} Following the same pattern as above, the solution sets at all other orders may be obtained by checking the linearized ``age-advection" equation, passing the solution for $\Delta_{::}$ into the linearized equation of state to obtain $\rho_{::}$, which may then be used to obtain $w_{::}$ from the linearized conservation of mass. Pressure $p_{::}$ is always obtained via integration of the $z$-component of linearized bulk momentum conservation equation after having determined $w_{::}$, and $u_{::}$ is always obtained by integration of the $x$-component of linearized bulk momentum conservation. 

\paragraph{At O($\epsilon^1 k^0 \hat{\alpha}^0$):}
We have
\begin{equation}
	w_{00}\tilde{\Delta}_{100}' = 0 \to \tilde{\Delta}_{100} = c_1
\end{equation}
and boundary condition at z=0, $\Delta_{00z}\tilde{g}_1 + \tilde{\Delta}_{100} = 0$, implies $\Delta_{100} = 0$, because $\Delta_{00z} = 0$. Then we immediately have $\rho_{100} = 0$. Mass conservation in the bulk yields
\begin{equation}
	\rho_{00}\tilde{w}_{100}' = 0 \to \tilde{w}_{100} = c_2
\end{equation}
and the boundary condition $\tilde{w}_{100} + w_{00z}\tilde{g}_1 = 0$ implies $\tilde{w}_{100} = 0$, because $w_{00z} = 0$. Then we have
\begin{equation}
	\tilde{p}_{100} = -2fA\eta D_{33}\{\tau_h \tilde{h}_1 + \tau_g \tilde{g}_1 \} + c_3
\end{equation}
by integrating the $z$-component of bulk momentum conservation. We also have steady-state stress tensor to leading order in $\hat{\alpha}$, $\textbf{T}_{00}$, whose bottom-right component is
\begin{equation}
	-p_{00} + 2\eta\{w_{00z} - fA\tau(z;0,h_0)D_{33}\}.
\end{equation}
With $p_{00} = -2fA\eta\tau(z;0,h_0)D_{33}$ and $w_{00} = V$ from previous calculations, $T_{00z}^{33} = 0$. This combined with the boundary conditions implies $c_3 = 0$. Then integrating the $x$-component of bulk momentum conservation twice and satisfying the boundary conditions leads to
\begin{equation}
	\tilde{u}_{100} = 2fAD_{13}\int_0^z \left( \tau_h\tilde{h}_1 + \tau_g\tilde{g}_1 \right) d\tilde{z} -2fAD_{13}\tau(0;0,h_0)\tilde{g}_1.
\end{equation}

\paragraph{At O($\epsilon^1 k^0 \hat{\alpha}^1)$:}
From linearized age-advection, we have
\begin{equation}
	\Delta_{101z} = \frac{1}{V}\frac{\partial \alpha_1}{\partial g}g_1 + \frac{\partial \alpha_1}{\partial h}h_1,
\end{equation}
and boundary condition at $z=0$, $\Delta_{01z}\tilde{g}_1 + \tilde{\Delta}_{101} = 0$ implies
\begin{equation}
	\Delta_{101} = \frac{1}{V}\int_{0}^z \left[ \frac{\partial \alpha_1}{\partial g}g_1 + \frac{\partial \alpha_1}{\partial h}h_1 \right]d\tilde{z} - \alpha_1(0;0,h_0)\frac{\tilde{g}_1}{V}.
\end{equation}
From the linearized equation of state, we have
\begin{equation}
	\tilde{\rho}_{101} = -\rho^* \tilde{\Delta}_{101},
\end{equation}
leading immediately to
\begin{equation}
	\tilde{\rho}_{101} = -\frac{\rho^*}{V}\left( \int_{0}^z \left[ \frac{\partial \alpha_1}{\partial g}g_1 + \frac{\partial \alpha_1}{\partial h}h_1 \right]d\tilde{z} - \alpha_1(0;0,h_0)\tilde{g}_1\right).
\end{equation}
With the x-component of linearized bulk momentum conservation, we have that $\tilde{u}_{101}$ is at most linear, and the homogeneous boundaries imply $\tilde{u}_{101} = 0$. Conservation of mass in the bulk leads to
\begin{equation}
	\rho_{00}\tilde{w}_{101z} + \tilde{\rho}_{101z}w_{00} = 0 \to \rho^*\tilde{w}_{101z} + V\tilde{\rho}_{101z} = 0,
\end{equation}
hence
\begin{equation}
	\tilde{w}_{101} = -\tilde{\rho}_{101}\frac{V}{\rho^*} + c_2,
\end{equation}
or
\begin{equation}
	\tilde{w}_{101} = \left( \int_{0}^z \left[ \frac{\partial \alpha_1}{\partial g}g_1 + \frac{\partial \alpha_1}{\partial h}h_1 \right]d\tilde{z} - \alpha_1(0;0,h_0)\tilde{g}_1\right) + c_2,
\end{equation}
and enforcement of the boundary condition at z=0
\begin{equation}
	\tilde{w}_{101} + w_{01z}g_1 = 0 \to \tilde{w}_{101} + \alpha_1(z;0,h_0)g_1 = 0
\end{equation}
implies $c_2 = 0$. Finally, from the integrated z-component of bulk momentum conservation, we have
\begin{equation}
	\tilde{p}_{101} = 2\eta \tilde{w}_{101z} + c_3.
\end{equation}
The bottom-right component of the steady-state stress tensor at O($\alpha$), $T_{01}^{33}$, is
\begin{equation}
	-p_{01} + 2\eta w_{01z} = 0,
\end{equation}
which implies $c_3 = 0$, hence
\begin{equation}
	\tilde{p}_{101} = 2\eta\left[\frac{\partial\alpha_1}{\partial g}\tilde{g}_1 + \frac{\partial\alpha_1}{\partial h}\tilde{h}_1 \right].
\end{equation}

\paragraph{At O($\epsilon^1 k^1 \hat{\alpha}^0$):}
From the linearized age-advection equation, we have
\begin{equation}
	w_{00}\tilde{\Delta}_{110}' = 0 \to \tilde{\Delta}_{110} = c_4,
\end{equation}
and the boundary conditions imply $c_4=0$, hence $\tilde{\Delta}_{110} = 0$. From the linearized equation of state, this implies $\tilde{\rho}_{110} = 0$, and linearized bulk mass conservation becomes
\begin{equation}
	i\rho^*\tilde{u}_{100} + \rho^*\tilde{w}_{110}' = 0 \to \tilde{w}_{110}' = -i\tilde{u}_{100}.
\end{equation}
Integration leads to
\begin{equation}
	\tilde{w}_{110} = -2ifAD_{13}\left[\int_0^z \int_0^{\tilde{z}}(\tau_h \tilde{h}_1 + \tau_g \tilde{g}_1 )d\tilde{z} d\tilde{\tilde{z}} - \tau(0;0,h_0)\tilde{g}_1z\right] + c_5,
\end{equation}
and boundary condition at $z=0$, $\tilde{w}_{110} = 0$ implies $c_5 = 0$. Then from the x-component of bulk momentum conservation, we have
\begin{equation}
	\begin{gathered}
		-i\tilde{p}_{100} + \eta \{ \tilde{u}_{110}''  + i\tilde{w}_{100}'\}= 2fA\eta i D_{11}\left[\tau_h\tilde{h}_1 + \tau_g\tilde{g}_1 \right]  \\ \to \\
		\tilde{u}_{110}'' = 2fAi(D_{11}-D_{33})(\tau_h\tilde{h}_1 + \tau_g\tilde{g}_1). 
	\end{gathered}
\end{equation}
Integrating twice leads to
\begin{equation}
	\tilde{u}_{110} = 2fAi(D_{11}-D_{33})\int_0^z\int_0^{\tilde{z}}(\tau_h\tilde{h}_1 + \tau_g\tilde{g}_1)d\tilde{\tilde{z}}d\tilde{z} + c_6z + c_7,
\end{equation}
and we must enforce boundary conditions. At z=0, $\tilde{u}_{110} = 0$, implying $c_7 = 0$, and at $z=h_0$,
\begin{equation}
	\begin{gathered}
		\eta\tilde{u}_{110}' + \eta i \tilde{w}_{100} - i\tilde{h}_1 T_{00}^{11} = 0 \\ \to \\
		\eta \left[2fAi(D_{11}-D_{33})\int_0^z(\tau_h \tilde{h}_1 + \tau_g \tilde{g}_1 )d\tilde{z} + c_6 \right] + 2fA\eta i \tilde{h}_1 \tau(z;0,h_0)(D_{11} - D_{33} ) = 0,
	\end{gathered}
\end{equation}
by recognizing that
\begin{equation}
	T_{00}^{11} = -p_{00} - 2fA\eta \tau(z;0,h_0)D_{11} = -2\eta fA\tau(z;0,h_0)(D_{11}-D_{33}).
\end{equation}
Then we find that
\begin{equation}
	c_6 = -2fAi\tilde{h}_1 \tau(h_0;0,h_0)(D_{11}-D_{33}) - 2fAi(D_{11}-D_{33})\int_0^{h_0}(\tau_h \tilde{h}_1 + \tau_g \tilde{g}_1 )d\tilde{z},
\end{equation}
which we will carry symbolically. Finally, we seek $\tilde{p}_{110}$. From the z-component of bulk momentum conservation, we have
\begin{equation}
	\begin{gathered}
		-\tilde{p}_{110}' + \eta(2\tilde{w}_{110}'' + i\tilde{u}_{100}) = 2fA\eta i D_{31}(\tau_h \tilde{h}_1 + \tau_g \tilde{g}_1) \\ \to \\
		-\tilde{p}_{110}' = 2fA\eta i(D_{31} + D_{13})(\tau_h \tilde{h}_1 + \tau_g \tilde{g}_1).
	\end{gathered}
\end{equation}
One integration leads to
\begin{equation}
	\tilde{p}_{110} = -2fA\eta i (D_{31} + D_{13})\int_0^z (\tau_h \tilde{h}_1 + \tau_g \tilde{g}_1)d\tilde{z} + c_7.
\end{equation}
From the boundary condition at $z=h_0$,
\begin{equation}
	-\tilde{p}_{110} + 2\eta \tilde{w}_{110}' = 0,
\end{equation}
hence
\begin{equation}
	c_7 = 2fA\eta i (D_{31} - D_{13})\int_0^{h_0}(\tau_h \tilde{h}_1 + \tau_g \tilde{g}_1)d\tilde{z} + 4fA\eta i D_{13} \tau(0;0,h_0)\tilde{g}_1,
\end{equation}
which we will carry symbolically. Last, we compute $\sigma_{10}$ from the compatibility condition due to the linearized kinematic condition at the upper interface,
\begin{equation}
	\begin{gathered}
		\sigma_{10}\tilde{h}_1 = \tilde{w}_{110} - u_{00}i\tilde{h}_1 + \frac{V\rho^*}{\rho_{00}^2}\tilde{\rho}_{110},
	\end{gathered}
\end{equation}
which leads to
\begin{equation}
	\sigma_{10} = -2fAi D_{13}\left[\int_0^{h_0}\int_0^{h_0}(\tau_h + \tau_g \frac{\tilde{g}_1}{\tilde{h}_1})d\tilde{z} - \tau(0;0,h_0)\frac{\tilde{g}_1}{\tilde{h}_{1}} h_0 \right] - 2fAi D_{13}\int_0^{h_0}\tau d\tilde{z}.
\end{equation}
Note the appearance of the interface relation term, $\frac{\tilde{g}_1}{\tilde{h}_1}$.

\paragraph{At O($\epsilon^1 k^1 \hat{\alpha}^1$):} 
Looking at the age-advection equation and simplifying, we obtain
\begin{equation}
	\sigma_{10}\tilde{\Delta}_{101} + iu_{00}\tilde{\Delta}_{101} + w_{00}\tilde{\Delta}_{111}' + \tilde{w}_{110}\Delta_{01z} = 0.
\end{equation}
Note the appearance of the interaction term $u_{00}\tilde{\Delta}_{101}$, which involves the depth-dependence profiles of both anisotropic plastic flow and isotropic swelling. This leads to
\begin{equation}
	\tilde{\Delta}_{111} = -\frac{1}{V}\int_0^z \left[\sigma_{10}\tilde{\Delta}_{101} + iu_{00}\tilde{\Delta}_{101} + \tilde{w}_{110}\frac{\alpha_1}{V} \right]|_{z=\tilde{z}} d\tilde{z} + c_8,
\end{equation}
and the boundary condition $\tilde{\Delta}_{111} = 0$ at $z=0$ implies that $c_8 = 0$. The linearized equation of state immediately reduces to
\begin{equation}
	\tilde{\rho}_{111} = -\rho^* \tilde{\Delta}_{111}.
\end{equation}
Bulk mass conservation leads to
\begin{equation}
	\begin{gathered}
		\sigma_{10}\tilde{\rho}_{101} + i\rho_{01}\tilde{u}_{100} + \rho_{01z}\tilde{w}_{110} + \rho_{00}\tilde{w}_{111}' + \rho_{01}\tilde{w}_{110}' + \tilde{\rho}_{111}'w_{00} = 0
	\end{gathered}
\end{equation}
after simplification. Then we have
\begin{equation}
	\begin{gathered}
		\tilde{w}_{111}' = \left(\frac{1}{V}\int_0^z \alpha_1 d\tilde{z} \right)\tilde{w}_{110}' + \frac{\alpha_1}{V}\tilde{w}_{110} + \left(\frac{i}{V}\int_0^z \alpha_1 d\tilde{z} \right)\tilde{u}_{100} - \sigma_{10}\frac{\tilde{\rho}_{101}}{\rho^*} - \frac{V}{\rho^*}\tilde{\rho}_{111}',
	\end{gathered}
\end{equation}
hence
\begin{equation}
	\tilde{w}_{111} = \int_0^z\left[ \left(\frac{1}{V}\int_0^z \alpha_1 d\tilde{z} \right)\tilde{w}_{110}' + \frac{\alpha_1}{V}\tilde{w}_{110} + \left(\frac{i}{V}\int_0^z \alpha_1 d\tilde{z} \right)\tilde{u}_{100} - \sigma_{10}\frac{\tilde{\rho}_{101}}{\rho^*} - \frac{V}{\rho^*}\tilde{\rho}_{111}'\right]|_{z=\tilde{z}} d\tilde{z} + c_9,
\end{equation}
and the boundary condition $\tilde{w}_{111} = 0$ at $z=0$ implies $c_9=0$. From the z-component of bulk momentum conservation, the knowledge that $\tilde{u}_{101} = 0$, and one integration, we have
\begin{equation}
	\tilde{p}_{111} = 2\eta \tilde{w}_{111}' + c_{10},
\end{equation}
and the boundary condition at $z=h_0$ implies $c_{10} = 0$. We now seek $\tilde{u}_{111}$ by looking at the x-component of bulk momentum conservation,
\begin{equation}
	\begin{gathered}
		-i\tilde{p}_{101} + \eta \{\tilde{u}_{111}'' + i\tilde{w}_{101}'  \} = 0 \\ \to \\
		\tilde{u}_{111}'' = \frac{i}{\eta}\tilde{p}_{101} - i\tilde{w}_{101}' = i\left[\frac{\partial \alpha_1}{\partial g}\tilde{g}_1 + \frac{\partial \alpha_1}{\partial h}\tilde{h}_1 \right],
	\end{gathered}
\end{equation}
where the final equality follows from simplification. Hence
\begin{equation}
	\tilde{u}_{111} = \int_0^z \int_0^{\tilde{z}}i\left[\frac{\partial \alpha_1}{\partial g}\tilde{g}_1 + \frac{\partial \alpha_1}{\partial h}\tilde{h}_1 \right] d\tilde{\tilde{z}}d\tilde{z} + c_{11}z + c_{12}.
\end{equation}
From the boundary condition $\tilde{u}_{111} = 0$ at $z=0$, $c_{12} = 0$. From the boundary condition at $z=h_0$,
\begin{equation}
	i\tilde{h}_1 p_{01} + \eta\left[\tilde{u}_{111}' + i\tilde{w}_{101} \right] = 0,
\end{equation}
we arrive at
\begin{equation}
	c_{11} = i\left[\alpha_1(0;0,h_0)\tilde{g}_1 - 2\alpha_1(h_0;0,h_0)\tilde{h}_1 - 2\int_0^{h_0}\left(\frac{\partial \alpha_1}{\partial g}\tilde{g}_1 + \frac{\partial \alpha_1}{\partial h}\tilde{h}_1 \right)d\tilde{z}\right],
\end{equation}
which we will carry symbolically. Finally, from the linearized kinematic condition and simplification, we obtain
\begin{equation}
	\sigma_{11} = \frac{1}{\tilde{h}_1}\left(\tilde{w}_{111} - V\tilde{\Delta}_{111} \right)|_{z=h_0}. 
\end{equation}

\paragraph{At O($\epsilon^1 k^2 \hat{\alpha}^0$):} 
From the linearized age-advection equation and the boundary condition at $z=0$, we find that $\tilde{\Delta}_{120} = 0$, which immediately implies that $\tilde{\rho}_{120} = 0$. Linearized bulk mass conservation and the boundary condition at $z=0$ for $\tilde{w}_{120}$ then imply
\begin{equation}
	\tilde{w}_{120} = -i\int_0^z \tilde{u}_{110}d\tilde{z}.
\end{equation}
From the z-component of bulk momentum conservation and simplification, we have
\begin{equation}
	\begin{gathered}
		-\tilde{p}_{120}' + \eta \{2\tilde{w}_{120}'' + i\tilde{u}_{110}'\} = 0 \\ \to \\
		\tilde{p}_{120}' = -i\eta \tilde{u}_{110}' \\ \to \\
		\tilde{p}_{120} = -i\eta \tilde{u}_{110} + c_{13}.
	\end{gathered}
\end{equation}
From the boundary condition at $z=h_0$, $-\tilde{p}_{120} + 2\eta \tilde{w}_{120}' = 0$,
we find that $c_{13} = -i\eta \tilde{u}_{110}|_{z=h_0}$, thus
\begin{equation}
	\tilde{p}_{120} = -i\eta \left[\tilde{u}_{110} + \tilde{u}_{110}|_{z=h_0} \right].
\end{equation}
Now we seek $\tilde{u}_{120}$. From the x-component of bulk momentum conservation, we have
\begin{equation}
	\begin{gathered}
		-i\tilde{p}_{110} + \eta\{-2\tilde{u}_{100} + \tilde{u}_{120}'' + i\tilde{w}_{110}'\} = 0 \\ \to \\
		\tilde{u}_{120} = \int_0^z\int_0^{\tilde{z}}\left[\frac{i}{\eta}\tilde{p}_{110} + 2\tilde{u}_{100} \right]d\tilde{\tilde{z}}d\tilde{z} - i\int_0^z\tilde{w}_{110} d\tilde{z} + c_{14}z + c_{15}.
	\end{gathered}
\end{equation}
From the boundary condition $\tilde{u}_{120} = 0$ at $z=0$, $c_{15} = 0$. From the boundary condition $\tilde{u}_{120}' + i\tilde{w}_{110} = 0$ at $z=h_0$, we have
\begin{equation}
	c_{14} = -\int_0^{h_0}\left[\frac{i}{\eta}\tilde{p}_{110} + 2\tilde{u}_{100} \right]d\tilde{z},
\end{equation}
which will be carried symbolically. Finally, using the linearized kinematic condition and the knowledge that $\tilde{\rho}_{120} = 0$, at $z=h_0$ we find
\begin{equation}
	\sigma_{20} = \frac{\tilde{w}_{120}}{\tilde{h}_1} =  -\frac{i}{\tilde{h}_1}\int_0^{h_0}\tilde{u}_{110}d\tilde{z}.
\end{equation}

\paragraph{At O($\epsilon^1 k^2 \hat{\alpha}^1$):}
We note that at this order, we require only $\tilde{w}_{121}$ and $\tilde{\rho}_{121}$ in order to determine $\sigma_{21}$ and complete our analysis. From the linearized age-advection equation with simplifications, we have
\begin{equation}
	\sigma_{10}\tilde{\Delta}_{111} + \sigma_{20}\tilde{\Delta}_{101} + w_{00}\tilde{\Delta}_{121}' + \tilde{w}_{120}\Delta_{01z} + iu_{00}\tilde{\Delta}_{111} = 0.
\end{equation}
Isolating $\tilde{\Delta}_{121}'$, integrating once and applying the boundary condition at $z=0$ implies
\begin{equation}
	\tilde{\Delta}_{121} = -\frac{1}{V}\int_0^z\left[\sigma_{10}\tilde{\Delta}_{111} + \sigma_{20}\tilde{\Delta}_{101} + \tilde{w}_{120}\Delta_{01z} + iu_{00}\tilde{\Delta}_{111} \right]d\tilde{z}.
\end{equation}
From the linearized equation of state and the knowledge that $\Delta_{00} = 0$ and $\tilde{\Delta}_{120} = 0$, we immediately find
\begin{equation}
	\tilde{\rho}_{121} = -\rho^*\tilde{\Delta}_{121}.
\end{equation}
Now we seek $\tilde{w}_{121}$ using the linearized bulk mass conservation equation. We find
\begin{equation}
	\sigma_{10}\tilde{\rho}_{111} + \sigma_{20}\tilde{\rho}_{101} + i\left[\rho_{00}\tilde{u}_{111} + \rho_{01}\tilde{u}_{110} \right] + \rho_{01z}\tilde{w}_{120} + \rho_{00}\tilde{w}_{121}' + \rho_{01}\tilde{w}_{120}' + \tilde{\rho}_{121}'w_{00} = 0.
\end{equation}
Rearrangement, integration and enforcement of the boundary condition for $\tilde{w}_{121}$ at $z=0$ implies
\begin{equation}
	\tilde{w}_{121} = -\frac{1}{\rho_{00}}\int_0^z\left[\sigma_{10}\tilde{\rho}_{111} + \sigma_{20}\tilde{\rho}_{101} + i\rho_{00}\tilde{u}_{111} + i\rho_{01}\tilde{u}_{110} + \rho_{01z}\tilde{w}_{120} + \rho_{01}\tilde{w}_{120}' + \tilde{\rho}_{121}'w_{00} \right]d\tilde{z}.
\end{equation}
Finally, we use the linearized kinematic condition at $z=h_0$ to determine $\sigma_{21}$ and complete the analysis. With $\tilde{\rho}_{120} = 0$, we have
\begin{equation}
	\sigma_{21} = \frac{1}{\tilde{h}_1}\left(\tilde{w}_{121} - V\tilde{\Delta}_{121} \right)|_{z=h_0}.
\end{equation}

\paragraph{Construction of general dispersion relation: general form.}
Recall that
\begin{equation}
	\sigma \approx 0 + k\left(\sigma_{10} + \hat{\alpha}\sigma_{11} \right) + k^2\left(\sigma_{20} + \hat{\alpha}\sigma_{21} \right). 
\end{equation}
Collecting all terms, we obtain a specialization of the expressions in the main text.

\section{Previous depth-dependence considerations}
In this part of the Appendix, we address some technical differences between the present work and that of \cite{moreno-barrado-etal-PRB-2015}.

\paragraph{Physical motivation and stress tensor form.} In both our work and that of \cite{moreno-barrado-etal-PRB-2015} treat the influence of the ion-beam as a viscous stress tensor with an extra term, essentially, 
\begin{equation}
	\textbf{T} = -p\textbf{I} + \eta\left(\nabla \vec{v} + (\nabla \vec{v})^T \right) + \textbf{T}_{\text{beam}}
\end{equation}
However, the form of that irradiation term is different between our respective analyses. In the present work, restricted to the $xz$ plane for the sake of comparison, we have
\begin{equation}
	\begin{gathered}
		\textbf{T}_{\text{beam,PW}} = -2fA\eta\tau(z;g,h) \times
		\begin{bmatrix} 
			\frac{3}{2}\cos(2\theta) - \frac{1}{2} & \frac{3}{2}\sin(2\theta) \\
			\frac{3}{2}\sin(2\theta) & -\frac{3}{2}\cos(2\theta) - \frac{1}{2} \\
		\end{bmatrix},
	\end{gathered}
\end{equation}
which has a fundamentally different physical meaning than that of \cite{moreno-barrado-etal-PRB-2015}. In their work
\begin{equation}
	\begin{gathered}
		\textbf{T}_{\text{beam,MB}} = \tau_{zz}(h)
		\begin{bmatrix} 
			\cos(2\theta) & \sin(2\theta) \\
			\sin(2\theta) & -\cos(2\theta) \\
		\end{bmatrix}.
	\end{gathered}
\end{equation}
The differences are subtle but impactful. First, $T_{beam,MB}$ modifies the amorphous layer by contributing stress, and the resulting flow fields are a response to this contribution, while $T_{beam,PW}$ is precisely the opposite: it deducts stress, permitting more strain per unit than would otherwise be possible. $T_{beam,PW}$ is motivated by previous work on the ``melt cycle" phenomenon in the electronic stopping regime, recently shown to lead to good experimental predictions even in the nuclear stopping regime, while $T_{beam,MB}$ appeals to analogies with classical hydrodynamic stability, seeking essentially to motivate a modification of the viscous flow comparable to that of the well-known case of flow down an inclined plane, hence their terminology of an ``effective body force". 

\paragraph{Experimental comparison; wafer-curvature measurements.}
The difference is not restricted to interpretation or motivation. We briefly consider differences in experimental predictions. The work of \cite{perkinsonthesis2017} has presented a first-of-its-kind angle-dependent measurement of ion-induced stresses, separating out the isotropic parts from the anisotropic parts. This permits an immediate and revealing comparison of our respective models. The analysis of \cite{moreno-barrado-etal-PRB-2015} results in an in-plane component of their steady-state stress tensor,
\begin{equation}
	T_{0,xx} = -p_0(z) + \tau_{zz}(0)\cos(2\theta),
\end{equation}
where
\begin{equation}
	p_0(z) = -\frac{\Delta\tau_{zz}}{R_0}z\cos^2(\theta) + \tau_{zz}(0)\cos(2\theta).
\end{equation}
As we have done in the present work, we may simply average across the film depth in order to compute $<T_{0,xx}>$, which can be compared with wafer-curvature measurements of Perkinson \cite{perkinsonthesis2017}. This leads to 
\begin{equation}
	<\tau_{0,xx}> = -\frac{1}{2}\Delta \tau_{zz} \cos^2(\theta)
\end{equation}
in the analysis of \cite{moreno-barrado-etal-PRB-2015}, compared with
\begin{equation}
	\begin{gathered}
		<T_{0}^{11}> = -6fA\eta\cos(2\theta) - 2\hat{\alpha}\eta.
	\end{gathered}
\end{equation}
in our own analysis. Revealingly, our stress tensor permits a change of sign for the in-plane steady-state stress component at some irradiation angle, while that of \cite{moreno-barrado-etal-PRB-2015} does not. The work of Perkinson \cite{perkinsonthesis2017} clearly shows such a transition of in-plane stress behavior around 50 degrees, consistent with our stress tensor \textit{even in the absence of isotropic swelling}, while that of \cite{moreno-barrado-etal-PRB-2015} \textit{is incapable of making such a prediction} unless $\Delta \tau_{zz}$ itself is allowed to change signs with angle-- which would be \textit{highly} unexpected in an ion-irradiated system, as the energy deposited in the rotating Gaussian ellipsoid model of Sigmund is generally a fair approximation of true energy deposition \cite{sigmund-PR-1969,sigmund-JMS-1973,hossain-etal-JAP-2012,hobler-etal-PRB-2016}.

\paragraph{Construction of depth-dependence: down-beam versus laboratory coordinates}
In \cite{moreno-barrado-etal-PRB-2015}, the depth-dependence about laboratory-coordinate $z$ is constructed by a finite difference approximation of the change in stress about the downbeam coordinate, $\partial_{z'}\tau_{zz} \approx \frac{\tau(h)-\tau(h_{ac})}{d_{z'}}$. However, in our opinion, coordinates in this construction seem to have been mixed: $h$ and $h_{ac}$ are treated as if they were in laboratory coordinates $(x,z)$ (indeed, they are later linearized), while $d_{z'}$ is a true down-beam distance \textit{in the downbeam coordinate}, a fixed quantity. If this is a finite difference approximation in the downbeam direction and the film is subject to the assumption that the ion-track deposits its power directly along the downbeam axis, leading to amorphization up to some lower interface, then the difference between the relevant quantities $\tau(h)$ and $\tau(h_{ac})$ along the downbeam direction is actually of fixed length, hence the symbols $h$ and $h_{ac}$ are fixed positions along the downbeam direction. As $h_{ac}$ is ``pinned" to $h$ due to the connection between the two via the downbeam ion-track, we would not expect them to be subject to linearization later in the analysis: the difference between them is fully characterized by the assumption of diagonally-translated interfaces and \textit{exact}. The downbeam film thickness $d_{z'}$ is likewise a fixed value and it is seemingly unnecessary to compute it in terms of local incidence angles. This renders the expansion $\cos(\theta-\gamma) \approx \cos(\theta) + h_{x}\sin(\theta)$ unnecessary in equations (8)-(13) of \cite{moreno-barrado-etal-PRB-2015}.

If $\tau(h)-\tau(h_{ac})$ is taken as a fixed quantity, no linearization of these terms is necessary; likewise, $\cos(\theta-\gamma) \approx \cos(\theta) + h_{x}\sin(\theta)$ is also unnecessary. Then equations (A15)-(A19), the equations of mass and momentum conservation in the bulk and the stress balance at the upper interface, are substantially modified. Specifically, all terms with $h_1$ due to the previously-discussed linearizations are set to zero. We note that the $h_1$ in (A20) still appears: it is due to the linearization of the lower interface, not due to the linearization of the stress tensor. However, with these $h_1$ terms gone from the bulk and stress balance at the upper interface, the term $\Delta\tau_{zz}$ completely disappears from the real part of the linear dispersion relation (although not from the imaginary part; an influence due to the lower interface is retained there, which is carried from the solution of the horizontal flow field at the lower interface). With these modifications, the real part of the linear dispersion relation is
\begin{equation}
	\text{Re}(\omega_q) = -\frac{48\tau_{zz}(0)R_0^2}{96\mu \sin^2(\theta)}\left(\sin^2(2\theta) + \cos(2\theta)\right),
\end{equation}
which predicts $\theta_c \approx 65^{\circ}$ independent of stress profile. The uniformity of this prediction is consistent with our own in the case of fixed downbeam film thickness, where we have assumed diagonally-translated interfaces as well.

\printbibliography
\end{document}